# *Nano-Nb Superconducting 3D Josephson Junctions Arrays in Cu-Normal Matrix: Superconducting Phase Coherence and Quantum Fluctuations Properties*

*R. L. de Almeida[1,2], J. Albino Aguiar[1], C. A. C. Passos[3], S. K. Sharma[4] and J. A. H. Coaquira[2]

[1]Department of Physics, Federal University of Pernambuco, 50670-901, Recife, PE, Brazil
[2]Núcleo de Física Aplicada, Institute of Physics, University of Brasília, 70904-970, Brasília, Federal District, Brazil
[3]Department of Physics, Federal University of Espírito Santo, 29075-910, Vitória, ES, Brazil
[4]Department of Physics, School of Basic Sciences, Central University of Punjab, Bathinda, Punjab 151401, India

*Corresponding author: e-mail: ralmeida@ifi.unicamp.br; almeida.almeida@gmail.com

## ABSTRACT

We investigate the superconducting behavior, phase coherence and quantum fluctuation properties of three-dimensional arrays of nanometric Nb Josephson junctions embedded in a normal Cu matrix (Cu–x wt% Nb, x = 3, 5, 15 and 20), with emphasis on the Cu10wt%Nb and Cu20wt%Nb compositions. Magnetic properties were measured by dc magnetization as a function of temperature and field in a SQUID magnetometer and by ac susceptibility. Transport properties, including magnetoresistance, *V(I)* curves and the critical current $I_{c(H,T)}$, were measured at fields up to 90 kOe. A micro-Hall sensor with ten transverse voltage bridges (5.0 × 0.2 × 0.01 $mm^3$) was used to calibrate the magnetic field actually applied to the sample. Magnetization and transport reveal a double superconducting transition that depends on the applied field and current: $T_{c(H)}$, with $T_{c(0)}$=8.2 K, associated with the intragranular (volumetric) response of the Nb grains, and a lower $T_{cJ(H)}$, associated with Josephson coupling between grains through the Cu matrix. The response combines that of bulk Nb with signatures of weakly coupled grains, consistent with a collection of superconducting grains in a non-superconducting matrix. The characteristic fields $H_{c1}$=400.03 Oe and $H_{c2}$ = 15 kOe (grains), and $H_{c1J}$ = 99.93 Oe and $H_{c2J}$=199.25 Oe (intergranular coupling), were determined. The critical current $I_{c(H,T)}$ shows a central maximum at low fields, followed by a long tail and small oscillations of period ΔH≈24 Oe, attributed to quantum interference between coupled grains. At higher fields, magnetic flux gradients destroy phase coherence and randomize the phase θ of the order parameter $\psi = |\psi| e^{i\theta}$, suppressing the oscillations. Magnetic and transport results are correlated and described by an effective-medium theory for superconducting grains in a normal matrix under magnetic field and current, supporting a three-dimensional Josephson junction array in a Cu matrix.

# I. INTRODUCTION

Since the discovery of superconductivity [1,2], the formation of a quantum condensate of paired electrons, described by a macroscopic order parameter $\psi = |\psi|e^{i\varphi}$, has been extensively studied in conventional (BCS) superconductors, which are classified as Type-I or Type-II.

There is a growing worldwide interest in superconducting systems at the nanoscale. The goals are to develop reliable methods for the controlled fabrication of materials and nanostructures of varying complexity for diverse applications, and to understand their physical properties, their relationship with structural characteristics, and their reproducibility through simulations [3–8]. Recent advances have been driven by significant investments in science and technology by developed nations, and the study of granular materials in particular shows great potential for technological applications [9–12].

Early studies of copper-rich niobium alloys [13] revealed the possibility of fabricating mechanically reinforced Cu–Nb normal conductors and Cu–$Nb_3Sn$ superconducting wires. The size of the Nb precipitates was found to depend strongly on the cooling rate [14] as the temperature decreases from the liquid to the peritectic reaction at $T \approx 1083$ °C [15]. The synthesis route used to obtain the samples studied here, and their structural characterization, were reported in a previous work [16].

A granular superconductor can be viewed as a collection of superconducting grains embedded in a matrix that may be a poor superconductor, a metal, or even an insulator. Granularity is therefore often invoked to explain transport properties that appear as two components. One is the intragranular contribution, which arises from grains with typical superconducting behavior. The other is the intergranular response, associated with weak-link superconductivity. The intragranular properties are intrinsic to the material, whereas the intergranular properties are extrinsic and depend on the sample conditions. The study of granularity predates the discovery of high-temperature superconductors (*HTSC*'s) [17]. For instance, in the 1970s, pressed pellets of low-temperature superconductors were used to introduce controlled defects in order to investigate the enhancement of the critical current through the addition of pinning centers.

Granularity reflects a breaking of symmetry at several length scales. At the atomic scale, it may result from atomic substitutions or localized defects such as impurities, vacancies and oxygen deficiency. At the nanoscale, it may arise from concentration gradients or secondary phases. At the micrometer scale, it may be due to extended defects such as grain boundaries and twin boundaries. In layered perovskites there is also a form of intrinsic granularity that follows from the layered structure itself [18]; it gives rise to the intrinsic Josephson effect, associated with current transport between weakly coupled $CuO_2$ planes. In single crystals and nearly perfect structures, granularity becomes a more subtle feature, which can be controlled through accessible parameters to induce and manipulate the related effects.

Networks of superconducting wires are another nanostructured system of interest. Such networks, fabricated by deposition of superconducting material on substrates through lithographic techniques, had their transport properties studied as a function of temperature and magnetic field by Pannetier and co-workers in the 1990s [19,20]. In 1994, Itzler et al. [21] measured the ac susceptibility of isotropic and anisotropic networks of superconducting aluminum, aiming to map the phase boundary between the superconducting and normal states from magnetic measurements and to compare it with the boundary determined from transport. Because such networks behave as arrays of Josephson junctions (JJs), whose phase dynamics is described by the Hamiltonian of Eq. (27) with the Josephson coupling energy $E_J = \frac{\Phi_0 I_J}{2\pi}$ (Eq. (29), Section III), their response is a useful reference for granular systems, which are considerably more complex. The results of Itzler et al. [21] show an evolution of the response with the applied magnetic field, and the problem deserves to be studied as a function of its parameters, such as the geometry of the network and the size of the platelets.

In this work we investigate the superconducting behavior, phase coherence and quantum fluctuation properties of nanometric Nb precipitates forming a three-dimensional array of Josephson junctions (3D-JJ) in a normal Cu matrix. We report dc and ac magnetization and magnetotransport measurements, including the critical current as a function of magnetic field, temperature and applied current, with a micro-Hall sensor used to determine the magnetic field actually applied to the samples. We observe a double superconducting transition: $T_c(H)$, with $T_c(0)=8.2$ K, associated with the intragranular response of the Nb grains, and a lower $T_{cJ}(H)$, associated with the coupling between grains through the Cu matrix. The critical current shows a central maximum at low fields, followed by small oscillations with a period $\Delta H_{period} \approx 24$ Oe, which are suppressed at higher fields when magnetic flux gradients destroy the phase coherence and randomize the phase $\phi$ of the order parameter $\psi = ¿\psi \vee e^{i\phi}$. The response combines that of bulk Nb with signatures of weakly coupled grains, which we interpret as Josephson coupling in a three-dimensional array of Nb particles, using an effective-medium theory for superconducting grains dispersed in a normal matrix under magnetic field and current [23].

# II. PROCEDURE AND EXPERIMENTAL DETAILS

## II.I. THE Cuxwt%Nb MICROSTRUCTURE EVOLUTION

To study the Josephson effect and related effects in weak links in the *CuXwt%Nb* system, we have prepared aconsiderable number de samples with x=3, 5, 10, 15 e 20, from materials with high purity Face-Centered Cubic lattice(Cu) ≈99.999% metal considered as“normal metal” for superconducting properties, and Body Centered Cubic Nb (BCC)≈9.99% with $T_{c-Bulk}^{(Nb)} = (9.25 \pm 2.02)K$ [1].The preparation of these samples is particularly difficult because high difference between the melting points of Cu. It's from Nb, and from the melting temperature during the cooling, there is a strong tendency of the formation of other phases between these two materials [14]. The first phase diagram constructed for Cu-Nb alloys [13] and the second, more complete, which includes two phase transformations independent of temperature, that is, a monothetic in approximately $T_{monothetic}^{(Cuxwt\%Nb)} \approx 1.545$ $^oC$ and a perithetic reaction in $T_{perithetic}^{(Cuxwt\%Nb)} = 1.083$ $^oC$ [15]. At this phase diagram, there are three phases dependent on temperature. The first phase $\alpha+\beta$, which lies in an isotherm immediately below $T_{perithetic}^{(Cuxwt\%Nb)}=1.083$ $^oC$ (perithetic region) consists of both constituents of the alloy in its solid phase. The phase $L+\beta$which is situated above T=1.083 $^o$Cand a non-isothermal line that goes up to the melting temperature of Nb at T≈ 2.468 $^o$C, depending on of the concentration *x* in the *Cuxwt%Nb* alloy, It is consisting of copper in the liquid phase and niobium in the solid phase. It is phase has a species of plateau (monothetic region) at $T_{monothetic}^{(Cuxwt\%Nb)}=1.545$ $^o$C. The third and final phase, the $L-phase$ which is situated above the line delimiting the phase $L+\beta$, and the region of the diagram where all two constituents of the material are in their liquid phase (depending on the concentration of Nb in Cu). The microstructure of the samples based on *Cuxwt%Nb* it's strongly dependent on the cooling rate $\frac{\partial T}{\partial t}$ of system [14], [24–26], and we verify through that, the faster the molten *CuXwt %Nb* system is brought to thermal equilibrium with the room temperature, better is the distribution and organized size of Nb particles in the Cu matrix. The solubility of Nb in Cu $\rightarrow S_{Nb \rightarrow Cu}$ is around $1\,wt\,\%$ at a temperature below $T_{1\,wt\,\%}^{(S_{Nb \rightarrow Cu})}$=1100 $^oC$, and this solubility decreases very rapidly with thereduction of temperature and at room temperature this same solubility $S_{Nb \rightarrow Cu}$ drops to a value less than $0.1\,wt\,\%$, $T_{0.1\,wt\,\%}^{(S_{Nb \rightarrow Cu})}$= 300 K [24–26]. In materials cooled in rates close to those achieved in this work, this solubility $S_{Nb \rightarrow Cu}$ does not exceed $0.15\,wt\,\%$. On the other hand, the border of the Nb is little known, however, there are experimental data [26] what indicate that the solubility of Cu in Nb $S_{Cu \rightarrow Nb}$ at temperatures close to $T_{0.07\,wt\,\%}^{(S_{Cu \rightarrow Nb})}$=1100 $^oC$ it's around $0.07\,wt\,\%$. To prepare the samples, the materials were weighed, and the concentrations x in samples with stoichiometry Cuxwt%Nb, was varied between x = 3, 5, 10, 15 e 20. The subsequent step was to melt the constituent materials (Cu) and (Nb) that were taken into the of an arc furnace chamber $I_{T=300\,K}^{max}$=500 A. After this step, we used a *RF* furnacein a "molten mass levitation configuration" to redistribute the Nb grains at the Cu matrix. A curious and interestingeffect we observe during the melting of the samples in the *RF* oven in an almost levitation configuration, it was the tendency of niobium to migrate to the peripheral regions of the samples. And there was a kind of involvement of all the melted mass by a thin Nb film. We label this effect as a "*cocoon effect*", because it was in fact a kind of Nb "*cocoon*" (will be reported elsewhere) that formed in around the melted mass during the melting of the samples in the *RF* furnace. We identified the sample regions where there was the highest Nb concentration, due to the diffusiveeffect during the process of *RF* fusion. For that, in a part of the sample labeled Cuxwt%Nb-11-02 (whose properties are mentioned in the table 1) we identify three regions, separate them and we focus on spot of the electron beam of the microscope scan for more information (chemical and compositional) on the microstructure of our samples.

Shown in Ref. [18]: Photo showing samples Cuxwt%Nb-11-02 (a) and the sample Cuxwt%Nb - 10-01 (b) immediately after they are melted in the arc furnace. If we look closely at the sample Cuxwt%Nb-11-02 (a) we perceive a typical brightness that is not characteristic of Cu (even this sample has x=80 % by weight of Cu). Microscopy images scanning revealed the migration of the Nb to the regions of the sample: "*cocoon effect*".

We estimate the cooling rate of the system and its limits, which take into account the downtime ( very short $\tau \approx 0.3\,s - 0.5\,s$) for the sample drop from the coil to the inside of the refrigerated Cu crucible based on the assumption that the emission of thermal energy through radiation by the sample and by heat transfer between the sample and the crucible to a temperature less than or equal to T=1000 $^oC$, where both elements are already found in the solid phase. We used the coefficient of Cu's thermal conductivity, which is 3.8 J/cm$^{2o}$C. We determine the rate in two stages: The first exclusively by radiation, taking in the emission of radiation between two bodies:

$$Q_{12} = F_{12}\,\sigma\,T_1^4\,A_1 \qquad (1)$$

where, $Q_{12}$ is the thermal energy from the body 1 (in if the sample flows into the entire inner chamber of the chamber, which is the body 2), $F_{12}$ is the superficial fraction of any and all the receiving area $A_1$ of energy seen by the area that is emitting (molten sample). It is valid to consider that for radials (which is the case with our system) $F_{12}=1$, T, and the temperature and $\sigma$ the Wien constant (or constant of Stephan-Boltzmann). And also, to estimate with greater (as much as possible) the rate of Fourier thermal conductivity equation

$$Q_{12} = -KA\left(\frac{\partial T}{\partial l}\right) \qquad (2)$$

Where K is the thermal diffusivity constant, in this case, Cu, which is the dominant quantity material in the sample, and The area effective contact between bodies. Overlapping equations 1 and 2 and with some workalgebraic we estimate the cooling rate (we will not reproduce the algebraic calculations here, we will show theresult) determined by the final expression:

$$\frac{\partial T}{\partial l} = \frac{\Delta T\, F_1 + \Delta T\, F_2}{\Delta l\, F_1 + \Delta l_{F2}} \approx 2800\, K/s \qquad (3)$$

Although (as has been mentioned) we have prepared a large number of samples Cuxwt%Nb ( 50), we will only discuss our results based on the samples whose properties (after are available in the table below. 1. After obtaining the samples by this last melting process, step was to cut the samples and prepare them for the structural characterization measures.

| **CuXwt%Nb Samples Properties** | | | | |
|---|---|---|---|---|
| Samples | Dimensions $l \times w \times t$ | Nb: **x wt%** | Mass (g) | Geometry |
| CuXwt%Nb - **10-01** | $0.50 \times 0.15 \times 0.016\ cm^3$ | $\mathbf{X = 10}$ | (0.0270±0.0001) | Rectangular |
| CuXwt%Nb - **11-02** | $0{,}40 \times 0,20 \times 0,01\ cm^3$ | $\mathbf{X = 20}$ | (0.0210±0.0001) | Retangular |

Table 1: Table containing information about sample properties CuXwt%Nb whose results of properties structural and superconductive properties are shown and discussed in this thesis. We add that the total number of samples obtained in this exceeded 50 prepared samples (all very good granular quality), however, in this text we will concentrate our discussions on the two samples mentioned in this table (Cuxwt%Nb-10-01 e Cuxwt%Nb-11-02).

In Ref. [16] are photos of the samples after cutting and polishing to the scanning microscopy. The sample Cuxwt%Nb-11-02 is shown inletter *a*,and sample Cuxwt%Nb-10-01 is shown in the letter *b* (both in Ref. [16]). We added that this cutting geometry was chosen only for the accomplishment of the structural characterization through the images of *Scanning Electron Microscopy*. For the measures of the superconducting properties (inductive and magneto-transport), the samples were prepared in the form of stripes, with dimensions at least five times smaller than the shown in Ref. [16] (see geometric dimensions of stripes, de *Cuxwt%Nb* in the table 1). In Ref. [16] we show the scanning electronmicroscopy images of both samples *Cuxwt %Nb*. As we can see, the geometry definition of Nb grains obtained in the copper matrix is much higherthanthat we had obtained the sub-micrometric grains of Nb, however, we still had the task of distribute them better throughout the volume of the Cu matrix.Thus, the last stage of sample preparation *Cuxwt%Nb*consisted of the use of a resistive furnace for a annealing of the sample system, with the objective of reaching only melting temperature of copper, $T = 1083\,℃$, and by the diffusion effects, since there is a density between Cu and Nb of $D_{Nb} - D_{Cu} = 0.56\, g/cm^3$.

Shown in Ref. [18]: Photo showing samples *Cuxwt%Nb*-11-02 (a) and *Cuxwt%Nb*-10-01 (b) ready tobe measured in the scanning electron microscope.

Shown in Ref. [18]: Electron microscopy images of sample scanning Cuxwt%Nb-10-01). As can be seen, the beam energy used for this image was $E_{el}=30\,kV$, the mode of obtaining the images was that of electrons backscatter. The grains of Nb in the Cu matrix dimensions smaller than $1\,\mu m$.

.

Thus, after a very smooth climb ramp temperature, $\frac{\Delta T}{\Delta t}=100\,℃/h$, we raise the system to an equal temperature a 1100 °C. We let the system stay in this temperature for one to two hours. After that, we reduced temperature on a faster ramp $\frac{\Delta T}{\Delta t}=250\,℃/h$. The results obtained with the rearrangement of the Nb grains in the Cu matrix can be seen in Ref. [16] for the sample Cuxwt%Nb-11-02. How can be observed in the figures, the distribution of Nb grains in the copper matrix is quite adequate, as there is a very good regularity in the spacing between the grains of Nb in the Cu matrix, and again, we did not find results similar to these in the literature. The geometry and granular spacing remain constant. At dark spots on some grains of Nb are due to propanone residues ($C_4H_8O$) what is it used to clean the surface of the samples prior to realization of the images, because, as discussed in chapter previous, the adopted measurement method was theelectron method backscattering , it's that in this way, it favors the elements with greater atomic number, andthe carbon, the main constituent of the ketone, has an atomic number smaller than the other two elements of the sample, and therefore the dark shadow color on some grains of Nb. From the results of the last annealing fusion, we made a statistical analysis of the sample Cuxwt%Nb- 11-02, using the scanning microscopy measurements, and we obtained theresults that are found in the graphic of the 5 and 6.

Shown in Ref. [18]: Electron microscopy image of scan showing the microstructure of the sample in more detail Cuxwt%Nb - 10-01. How can be the energy of the beam used for this image was$E_{el}=30\,kV$, the mode of obtaining the images was that of electrons backscatter. The grains of Nb in the Cu matrix have smaller dimensions than $1\,\mu m$.

Shown in Ref. [18]: Electron microscopy images of scan showing the sample microstructure Cuxwt%Nb – witha better distribution of grains along the matrix.

Through our analysis we obtainedthe following microstructural results found in the graph of Ref. [16]:

- $d_m 1.2\,\mu m$, where $d_m$is the distance between grains.
- $D_m > 2.2\,\mu m$, where $D_m >$¿is the mean diameter of Nb grains.
- $\rho\,1.785\,grain/\mu m^2$, where is the mean grain density of Nb in matrix of Cu.

Shown in Ref. [18]: Electron microscopy image of with the microstructure of the sample Cuxwt%Nb-11-02 in details.

After completing the first stage, which consisted of preparation (growth in three merger processes), microstructural characterization by Scanning Electron Microscopy-(SEM-JEOL$^{Ltd}$, qualitative and quantitative*EDX* and *EDS* will be showed elsewhere) as shown above, we will show the physical characterization properties for the samples Cuxwt%Nb (x=3, 5, 10, 15 e 20), with focus for the concentrations x=10% and 20%. We will show below theresults of$M^{dc}$[H,T,$\theta$], $\chi^{ac}[T, H, h_{Exc}, \upsilon, \theta$, $\rho^{dc}$[H,T,$\theta$], $R^{dc}$[V,IH,T,$\theta$], I[V,H,T,$\theta$] and J[H,T,I,$\theta$]characterization of the superconducting properties of these granular samples.

## II.II.Cuxwt%Nb MAGNETIC MEASUREMENTS: $\chi^{ac}[T, H, h_{Exc}, \upsilon, \theta$AND $M^{dc}$[H,T,$\theta$]

The coherencedenoted as ξ of the superconducting state in type II Low temperature superconductivity-LTSC's (also in type I) is a manifestation of the quantum character macroscopic condition of this state. While the coherence of the electronic waveforms in the normal state is limited in the microscopic or macroscopic scales of length in the presence of spreading centers (usually inelastic scattering), which in practice are major impediments to the properties of transport, electrons in the superconducting state (as we have said) can be described by a macroscopic wave function (the superconducting order parameter) $\psi = |\psi|e^{i\phi}$which is a function essentially undisturbed by origin scattering. magnetic field. In the absence of magnetic fields and electric currents, the phase amplitudes of $\psi$are uniform in space in large macroscopic distances. In the presence of electric currents, a finite phase gradient, but coherence is (the phase of the order parameter remains unchanged, random and its mean is not zero). This statementcorroborates that we see in practice, because,as we know, there are superconductors in the construction of current transport wires electrical power, and these reach several kilometers maintaining its electrical resistance null. The vortices lattice that arises under strong applied magnetic fields in the mixed state, does not destroy its coherence, since the still not random because of the long-range ordering of the vortex lattice. Coherenceis preserved at the borders of the sample, since the amplitude$\psi$ is the same amplitude of the bulk and the phase is still non-random[27–29]. And is the preservation of this coherence at the boundaries of the material (superconducting grains) that allows us to observe of the effects treated in this work, such as the detection of the Josephson effect[30] and its applications [31]. The BCS theory that explains very well the microscopic coherenceorigin. For when the electrons are in the state paired (forming the Cooper pairs) into a LTSC, there are many Cooper pairs within the radius of a single pair that has valor ξ[32]. Intuitively, it is possible to infer that in a given region of the space, it is not possible to destroy only a single pair of Cooper, but many. And this will cost a significant amount of energy, basically, an amount proportional to the volume of condensation per unit of volume coherence, which is denoted by $\xi^3$. Thus, we can estimate the typical value of this energy in a LTSC (in this case, Nb, which is the LTSC superconducting element of our granular samples). We know that the number of electrons condensed per unit volume is given strictly by the density of energy states occupied in the normal state$N(0)$multiplied by BCS gap Δ, seeJ. Bardeen, L.N. Cooper, J. R. Schrieffer in [32]and L. P. Gorkov [33]). Thus, the energy needed to destroy all the Cooper pairs in a volume of coherence $\xi^3$is given by:

$$U \approx n \frac{\Delta}{E_F} x \xi^3 \times \Delta \qquad (4)$$

here we call n of electronic density of states using the relationship$N(0) n = E_F$, where $E_F$is the energy Fermi level energy . If we assume typical values for a LTSC (Nb case in those samples), or $n = 10^{22}/cm^3$ Δ= $1\, meV$, $E_F = 5\, eV$ and $\xi = 1000\, \AA$ . Thus, we get a value of the order of

$$U_{LTSC}^{Cuxwt\%Nb} \approx n \frac{\Delta}{E_F} \times \xi^3 \times \Delta = \frac{10^{22}}{cm^3} \times \frac{1\, meV}{5\, eV} \times 1000^3 \times 1\, meV \approx 1.0\, eV \quad (5)$$

We immediately realized that this energy is very large, particularly when compared with thermal energy, $k_B T$ at relevant temperatures to the LTSC,$U_{LTSC}^{Cuxwt\%Nb} \gg \Delta$. So we can understand this result as a confirmationthat there are many Cooper pairs of within a single pair radius, and that the energy necessary to destroy the superconductive coherence in a volume consistency is much greater than the energy of the gap. What else important we can draw from this result for the system that in this text is that, in our view, imperfections (or singularities) such as defects, grain boundaries, dislocations and defects on atomic scale have no impact whatsoever on the consistency of superconductivity in the samples *Cuxwt%Nb* studied in this work (where Nb is a Type II LTSC). In the figures 1 e 2 we show four curves of magnetization as a function of temperature for the sample Cux10%Nb-10-01, as can be seen immediately, the curves have two clear transition temperatures,$T_C^{M\times T}$and $T_{CJ}^{M\times T}(H)$. Increasing the appliedfieldto the temperature of these transitions decreases. For High-temperature superconductors (HTSC) samples, this transition was explained in some papers [34,35] as due to the Josephson region coupling between regions (or grains) of higher $T_c$, distributed by the volume of the sample. These results a common feature, which$H_{C2}\times T$is accepted as an intrinsic property of samples. And the line$H_C\times T_{CJ}$is based on the Josephson coupling between the superconducting grains dispersed in a normal matrix, which is an extrinsic property of the samples. That is, the superconducting properties of these samples are strongly dependent on the conditions of preparation. In this superconducting system Nb grains dispersed in a matrix normal, we can study more detailed aspects of this line $H_C\times T_{CJ}^{M\times T}(H)$, using the extrinsic property of the system, and control over the granularity of the samples. From the$M^{dc}$[H,T,$\theta$] measurements at thefigures 7 e 8, we can extractthe very important following information:

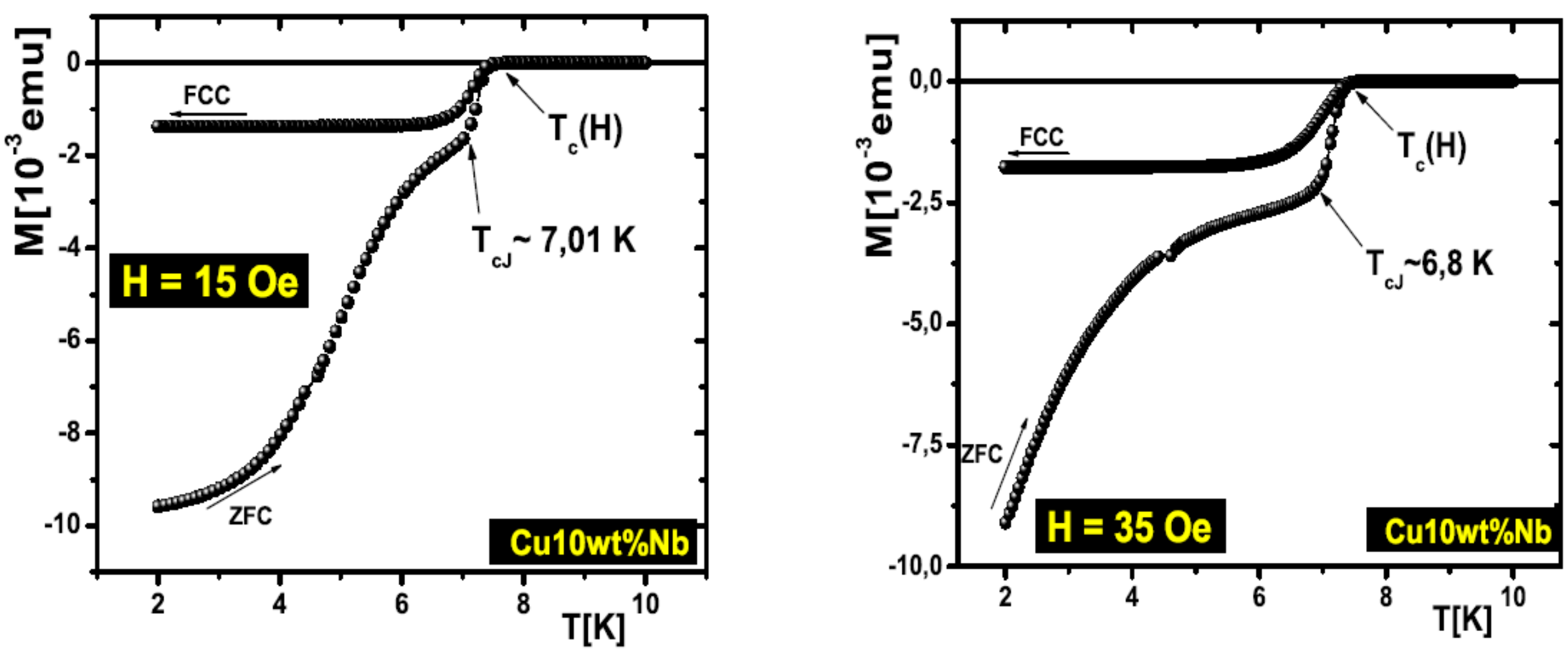

Figure 1: Measures of magnetization as a function of temperature for the sample Cux10%Nb-10-01. The bulk transition temperature superconducting is labelled as$T_C(H)$ and the transition temperature due to the weak coupling is labelled as$T_{CJ}^{M\times T}(H)$: a) measure taken in H=15 Oe and b) Measure fulfilled H=35 Oe. The granularCuxwt %Nb superconducting system properties have peculiar extrinsic characteristics, where the clusters (grains) superconductors and ways of coherence through non-randomness of the phase of the superconducting order parameter $\Psi$, modulated by the Josephson coupling. What is a behavior often related to weak coupling due to the existence ofsuperconducting weak links between the grains (regions).An important way to control the temperature value of transition $T_{CJ}^{M\times T}(H)$is the variation of the granular parameters of samples, that is, the control of the concentration of Cu in samples Cuxwt%Nb. This can be done by varying the amount of Nb in *wt* % or *at*% during the preparation process of these samples, as describedin detailsat the phase diagram discussion. And always important to be aware of two factors that we consider very important. The first one of these is the solubility (or miscibility) of Nb in Cu $S_{Nb \to Cu}$and the solubilityCu in Nb$S_{Cu \to Nb}$, therefore, these values ($S_{Nb \to Cu}$ and $S_{Cu \to Nb}$) can change the nominal parameters of the samples and can exert influence on the granularity of the system. Another important factor, completely new that we detected in the preparation of these samples, was the effect we call the "*cocoon effect*", that is, the migration of Nb in large quantities for the peripheral regions of the sample (see Ref. [16]). Of course, the "*cocoon effect*" is determinant in the effective granular properties of the samples *Cu*x*wt%Nb*, because there is migration of large amounts of Nb to the outside the melted*Cu*x*wt%Nb* mass amount. In accordance with our experiments in the preparation of these samples, this is a parameter of very difficult that determines the good granularity and homogeneity of the system *Cu*x*wt%Nb*. In fact, granularity is the key parameter experimentally controllable in these samples. For samples that are granular in the macroscopic sense, the granular fraction ends being a measure of the relative superconducting volume which, in principle, canbe controlled during preparation.

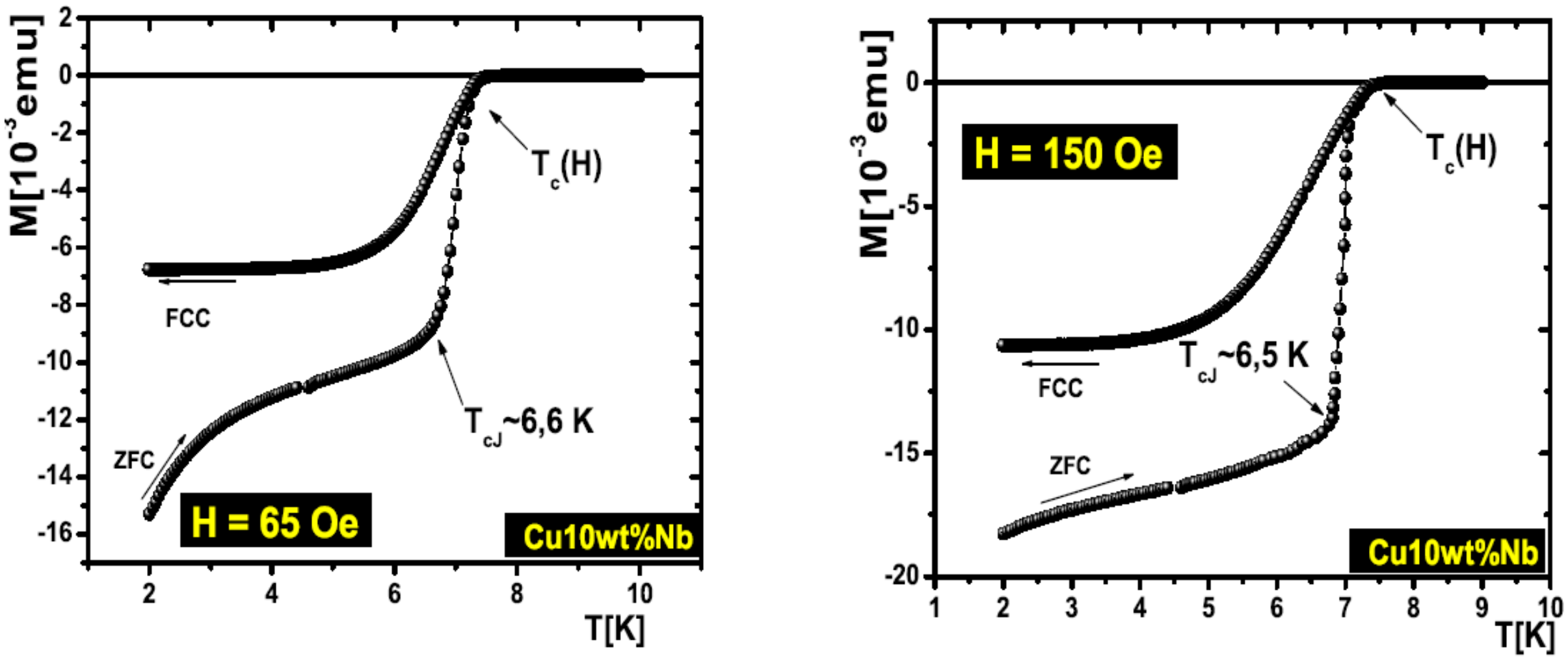

Figure 2: Second series of temperature-dependent magnetization measurements for sample Cux10%Nb-10-01. The bulk transition temperature superconducting is labelled as$T_C(H)$and the temperature of transition due to weak coupling is labelled as$T_{CJ}^{M \times T}(H)$: a) measure taken in H = 65 Oe b)measure taken in H = 150 Oe.

However, for single crystals and other almost perfect structures, granularity is a much more subtle feature, but can also be controlled in such a way that effects on it can be induced and tuned through parameters accessible[36]. Thus, we understand that for concentrations of Nb in wt % less than x=10, the effective granularity of the system is strongly affected and differs completely from the nominal concentration. In the graphics of the figures 3, 4, 5 and 6, we show the ac magnetic susceptibility measurements $\chi^{ac}[T, H, h_{Exc}, \upsilon, \theta$ for the sample Cux20%Nb-11-02.

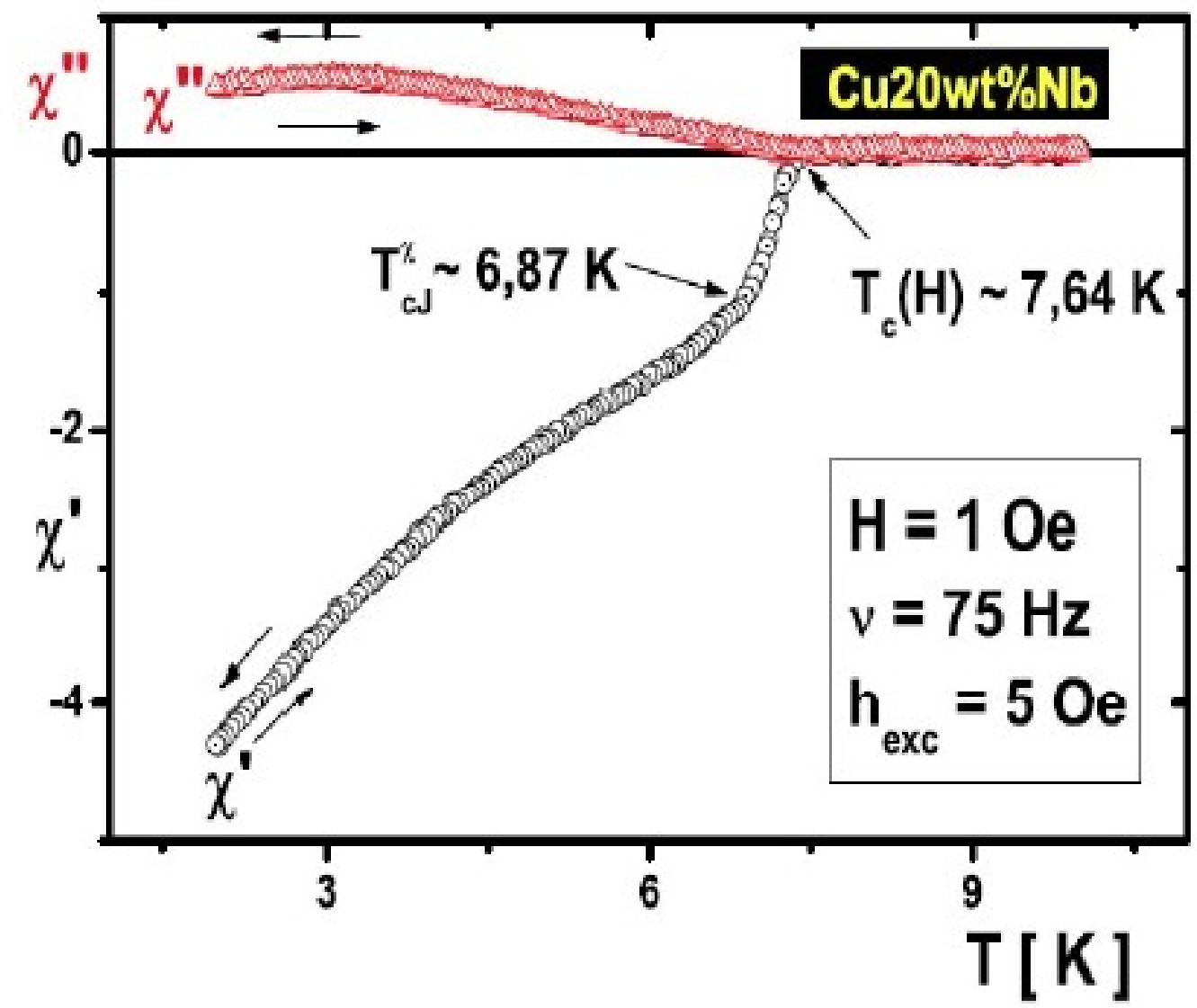

Figure 3: $\chi^{ac}[T, H, h_{Exc}, \upsilon, \theta$susceptibility measures for the sample Cu20wt%Nb-11-02. Measure taken in $H^{appl}=1\,Oe$, $\upsilon=75\,Hz$and$h_{Exc}=5\,Oe$.We found a strong dependence on temperature$T_{CJ}^{M\times T}(H)$ with the applied magnetic field. Thesuperconducting transition bulk is labelled as$T_C(H)$and the transition temperature dueto the weak coupling is labelled as $T_{CJ}^{\chi}$.

The experimental ac susceptibility measure was madeinductively using static magnetic fields (dc)$H^{appl}$ or alternating (ac)$h_{Exc}$ or both ($H^{appl}+h_{Exc}$). In these measures itis possible (by configuration of the PPMS platform), eliminate the uniform signal produced by the excitationfield and also part of the noise generated by external sources, leaving only thereal sample magnetic signal. Like this, when we measure the induced voltage in the secondary circuit, only the signal whose frequency is equal to or equal to one frequency multiple of the excitation ac field $h_{Exc}$. In this way, any noise that has a different frequency than that used for the excitation fielddoes not interfere with measurement signal of the ample. This frequency selection, together with the careful calibration of the secondary circuit, transforms the ac be an extremely sensitive technique. Granular superconductors samplespresent as its main property the physical connection of the superconducting grains inside the matrix. And These grains have a diamagnetic response corresponding to the sum of the individual responses of all grains. However, a collective response overlaps with the individual, so that the system displays a two-component response, one inter-granular and one intra-granular as can be seen in the curves 7,8,9,10,11 and 12. The intra-granular component $T_C(H)$is considered inherent to the material (Nb with some Cu solubility inside as discussed), and associated with the inter-granular $T_{CJ}^{\chi}$material has which is manifested in the physical properties of the sample: Josephson Coupling like. Thus, these superconducting samples have something remarkable, is that the largest can be carried through the sample *e.g.*, its critical current density $J_C$, is determined by the critical inter-granular current. This is not surprising, since that the transmission of current is done through a succession of grains bound in series by inter-granular material. The “critical point” of this series $J_C$association is, of course, limited by its weakest link. In the measurements of the graphs 9,10,11 and 12, the real component of susceptibility$\chi^{ac}[T, H, h_{Exc}, \upsilon, \theta$ decreases in magnitude (it becomes less diamagnetic) with the external dc magnetic field increase ($H^{appl}=1\,Oe, H^{appl}=2\,Oe$, $H^{appl}=3\,Oe$ and $H^{appl}=5\,Oe$), which means that there is a lower intensity of flow excluded. Perhaps because of very dissipative thermal and as well as effects of demagnetization, we did not observe the same time the expected response, which is the other component, the complex susceptibility$\chi''$ pass a maximum, but even then, the behavior shown per$\chi''$, indicates that there is dissipative movement of flow in the material. For the situation represented in the figures 3,4,5 and 6, between$T_{CJ}^{\chi}(H)=¿$6.36 K - $T_{CJ}^{\chi}(H)=¿$6.87 K, there is a change in the rate of growth of $\chi'$, which remains more pronounced at temperatures in between$T_C(H)=¿$ 7.31K - $T_C(H)=¿$ 7.64 K .

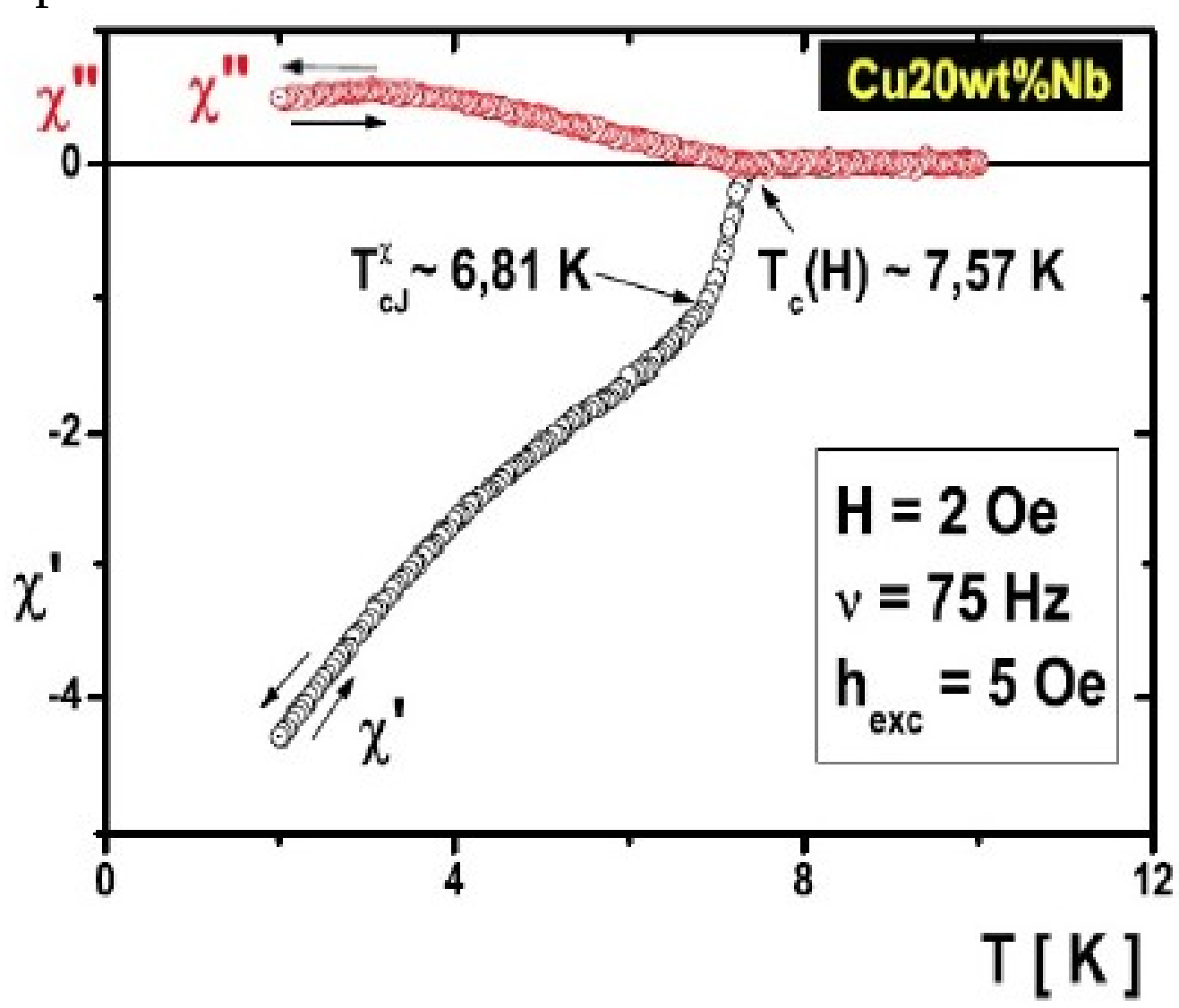

Figure 4: $\chi^{ac}[T, H, h_{Exc}, \upsilon, \theta$susceptibility measures for the sample Cu20wt%Nb-11-02. due to the weak coupling is labelled as $T_{CJ}^{\chi}$. Measure taken in $H^{appl}=2\,Oe$, $\upsilon=75\,Hz$ and$h_{Exc}=5\,Oe$.

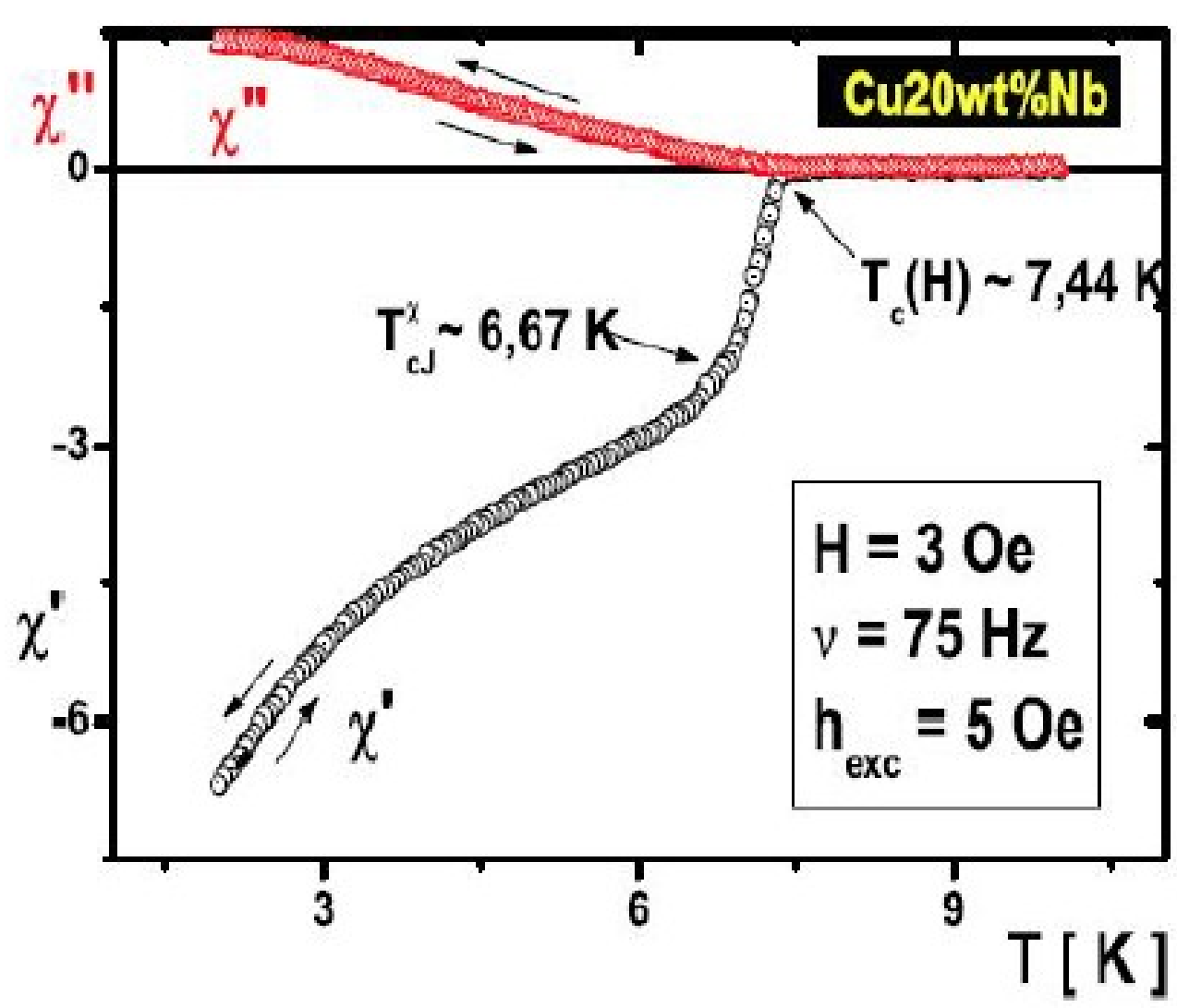

Figure 5: Second series of$\chi^{ac}[T, H, h_{Exc}, \upsilon, \theta$susceptibility measures for the sample Cu20wt%Nb-11-02 to find $T_{CJ}^{\chi}$. Measure taken in $H^{appl}=3\,Oe$, $\upsilon=75\,Hz$ and$h_{Exc}=5\,Oe$.

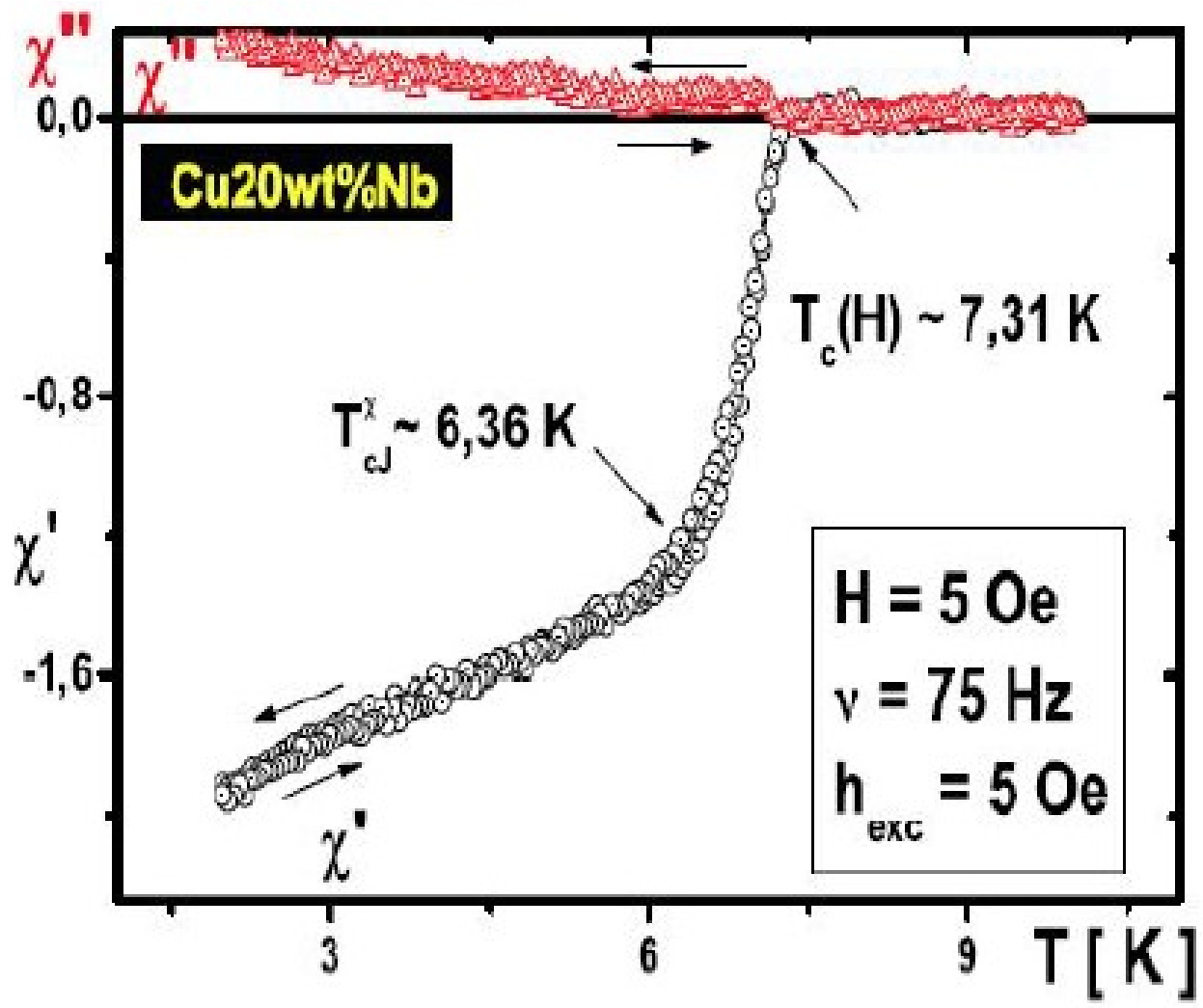

Figure 6: $\chi^{ac}[T, H, h_{Exc}, \upsilon, \theta$ac susceptibility measures for the sample Cu20wt%Nb-11-02 to find$T_{CJ}^{\chi}$. Measure taken in $H^{appl}=5\,Oe$, $\upsilon=75\,Hz$ and$h_{Exc}=5\,Oe$.

At the subsequent interval of temperatures, $T>7.64\,K$ what happens is usually called as second patamar, though$\chi^{'}$. Do not be properly horizontal. The main reason for the name is that, in fact, the slope would be zero for a sample that exhibited properties completely homogeneous in inter-granular surfaces, and the curve would exhibit a "*second floor*". Thus, at low temperatures the magnetic flux is excluded from the whole sample, including the matrix inter-granular, and we have $\chi^{'}-1$and $\chi^{''}0$. By increasing the temperature, it is expected that the inter-granularcritical current is achieved slowly in parts of the sample, given the inhomogeneity, this time from the Cu presence,and we have a decrease in the absolute value of $\chi^{'}$ (smaller capacity of the field) accompanied by an increase in the dissipative activity associated with flow penetration. Eventually, the temperature reaches sufficiently high values to overcome the critical grain current, which admit fieldin its interior. This shape of the ac susceptibility, characteristic of granular superconductors, is known in the literature as *two peaks - two drops* (*two peaks – two falls*)[36]. For during the penetration of flow inside the grains$\chi^{'}$and $\chi^{''}$reproduce the same inter-granular penetration. And the dissipative activity during the flow penetration inside the grains is considerably less intense than in the corresponding process. of inter-granular penetration. Returning to the graphics of the figures 1-6, toemphasize that there is dependence on $T_{CJ}^{\chi}$, or dependency $T_{CJ}^{MXT}(H)$ in the measures with the fieldapplied. And this is the inherent characteristic ofthese systems, that is, the strong and fundamental dependence of systems whose superconducting propertiesare governed by the Josephson Effect. It is important to consider that because the susceptibility to be a measureinvolving several thermodynamics parameters, we take careto keep them constant, this is,$\upsilon=75\,Hz$ and $h_{Exc}=5\,Oe$, except for the$H^{appl}$field, which wasvaried for each measure. Now, let's discuss the $H\times T$magnetic phase diagram $H\times T$for the samplesCux20%Nb and Cux10%Nb. At the figure 7, we show the magnetic phase diagram $H\times T$ obtained for thesample Cu10wt%Nb from the $M\times T$ measurements. The critical lines $H_C$ and $H_{CJ}$are defined at the transitiononset and the weak coupling transition where were extracted the $T_{CJ}^{MXT}$data points. At the lower inset we showthe region of magnetic fields $H_{C1J}\leq H_{critical}^{lower}\leq H_{C2J}$that delimitates the region that is controlled by the Nb grains weak coupling. As can be seen in this curve at the figure 7 and 8, there is clearly a system “inserted within the other”, which are represented by their curves-diagrams $H \times T$ and $H_{CJ} \times T$. From the measurement shown in figure 9, It is possible to clearly identify the presence of two regions for the grains of Nb in the sample Cu20wt%Nb. The dominated behavior by Josephson coupling, whose magnetic response is in the region delimited by $H_{C1J} \leq H \leq H_{C2J}$, it is exclusively granular and the other region of response behavior exclusively volumetric grains of Nb, which is delimited by $H_{C1} \leq H \leq H_{C2}$. The region of fields situated in $H \leq H_{C1J}$ is a region that we are sure is completely Meissner. Similarly, to what was done for the measure shown in graphic of figure 9, for the sample Cu20wt%Nb, we can also identify the regions mentioned in the graph 25 of magnetization as a function of the magnetic fieldmeasured in the sample Cu10wt%Nb. We also identify here in Ref. [XX], two regions for the Nb grains, behavior dominated by the Josephson coupling, whose magnetic response lies in the region delimited by $H_{C1J} \leq H \leq H_{C2J}$. It is exclusively the granular region and the other region of volumetric response of the grains, which is delimited by $H_{C1} \leq H \leq H_{C2}$. Also, is shown the region of fields located in $H \leq H_{C1J}$ the regime completely Meissner. As is shown in the figure 12 it is possible to quantify the regions identified in the measure shown in the graph 15. We measured for the sample Cux20%Nb-11-02 at T = 2K the following values o $H_{C1J}$, $H_{C2J}$, $H_{C1}$, $H_{C2}$ and $T_C$: $H_{C1J} = 99.93\,Oe$, $H_{C2J} = 199.25\,Oe$, $H_{C1} = 400.03\,Oe$ and $H_{C2} = 1.5\,T$, remembering that the measured bulk critical temperature this sample is $T_C$=8.2K. We can understand this curve, figure 9, of the following way: For magnetic fields $H \leq H_{C1J} = 99.93\,Oe$ the answer is exclusively Meissner, however, for magnetic fields in the interval $H_{C1J}(99.93\,Oe) \leq H \leq H_{C2J}(199.25\,Oe)$ the magnetization response is completely due to the coupling of Nb grains in the matrix, for an equal magnetic fieldvalue than $H_{C2J} = 199.25\,Oe$ but smaller than $H_{C1} = 400.03\,Oe$, the frustration of the coupling between the grains dominate due to the presence of the pressure of the magnetic field, which in addition to the grains, breaking.

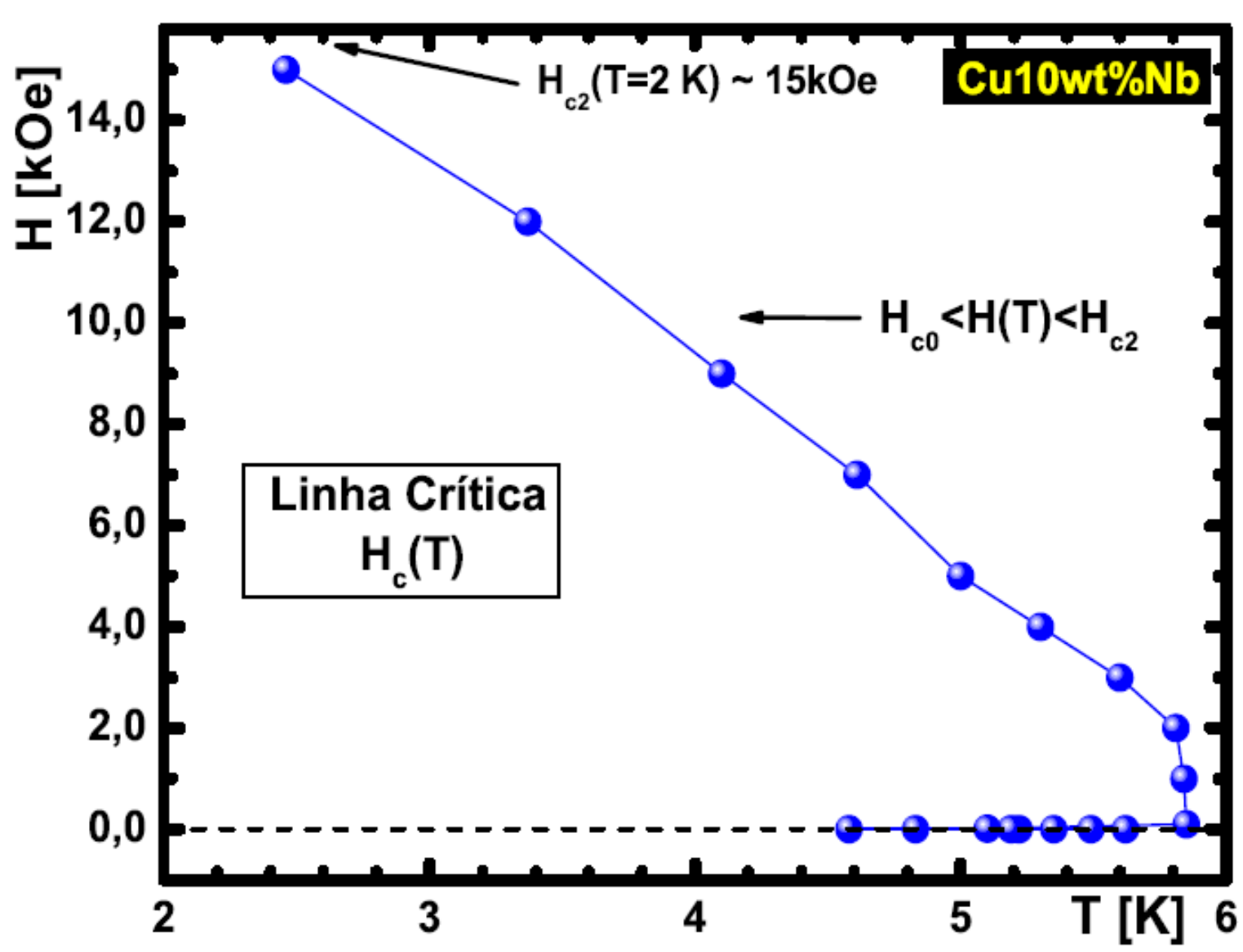

Figure 7: Phases diagram $H \times T$ built for the sample Cu10wt%Nb The from the $M \times T$ measurements. The Critical lines $H_C$ and $H_{CJ}$ are defined in the onset of the transition and in the transition from weak coupling from which points were extracted $T_{CJ}^{MXT}$. At the lower inset we show the region of fields magnetic $H_{C1J} \leq H_{critical}^{lower} \leq H_{C2J}$, that delimits the region governed by weak coupling between the Nb grain.

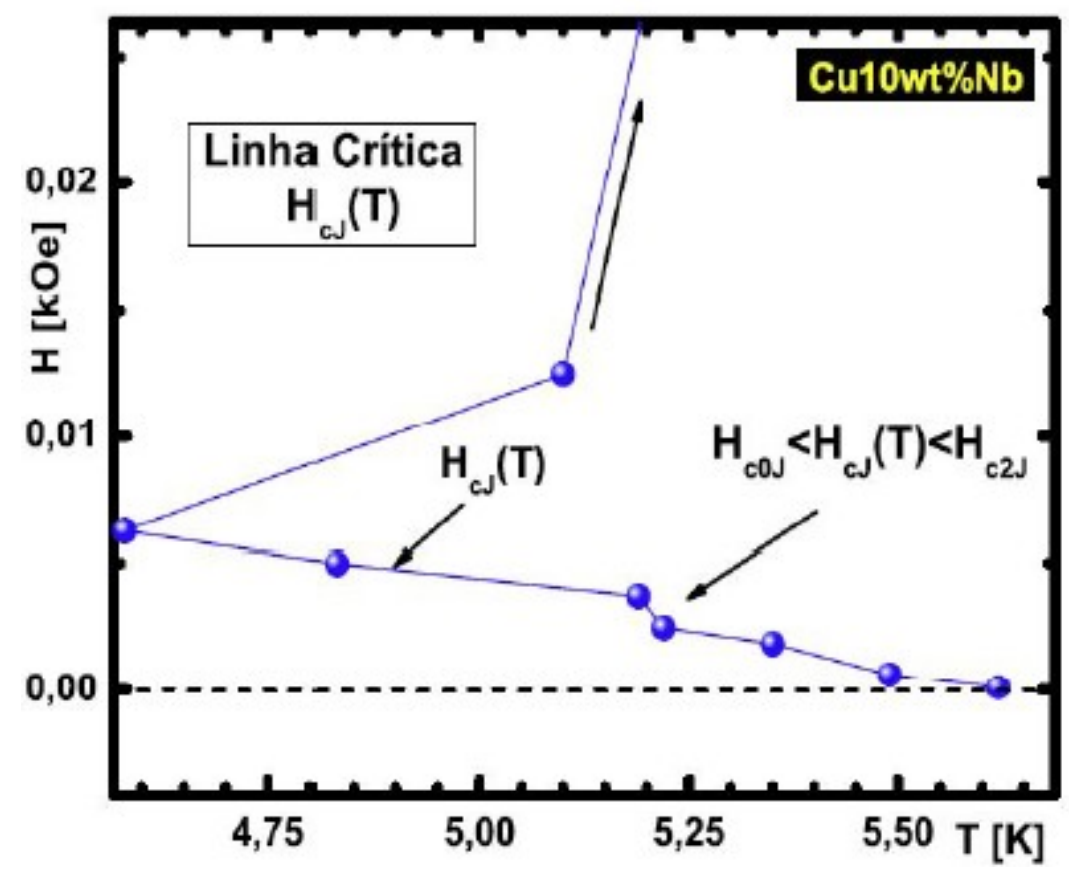

Figure 8: Inset built for the sample Cu10wt%Nb from the enhancement of the phase diagram $H \times T$ showed at the figure 7 in the temperature region $T_{inf}(4.55\,K) \leq T \leq T_{\square}(5.75\,K)$.

The Cooper pairs coherence stops and are no more interfering constructively between any grains. Thus, in $H_{C2J}$ the answer due to thegranular coupling ceases to exist. The from $H_{C1}$ penetration of the magnetic flux into the grains in the form of vortices, and this process of flow penetration increase with the applied external magnetic field $H^{appl}$. In the vicinity of $H_{C2}$ the density of flow lines reaches its critical value, thus, normal nuclei of these flux lines begin tocollapse and the sample collapse and goes through a phase transition of second order exactly in $H_{C2} = 1.5\,T$, and the system from $H_{C2}$ responds with normal behavior related their superconductive properties.

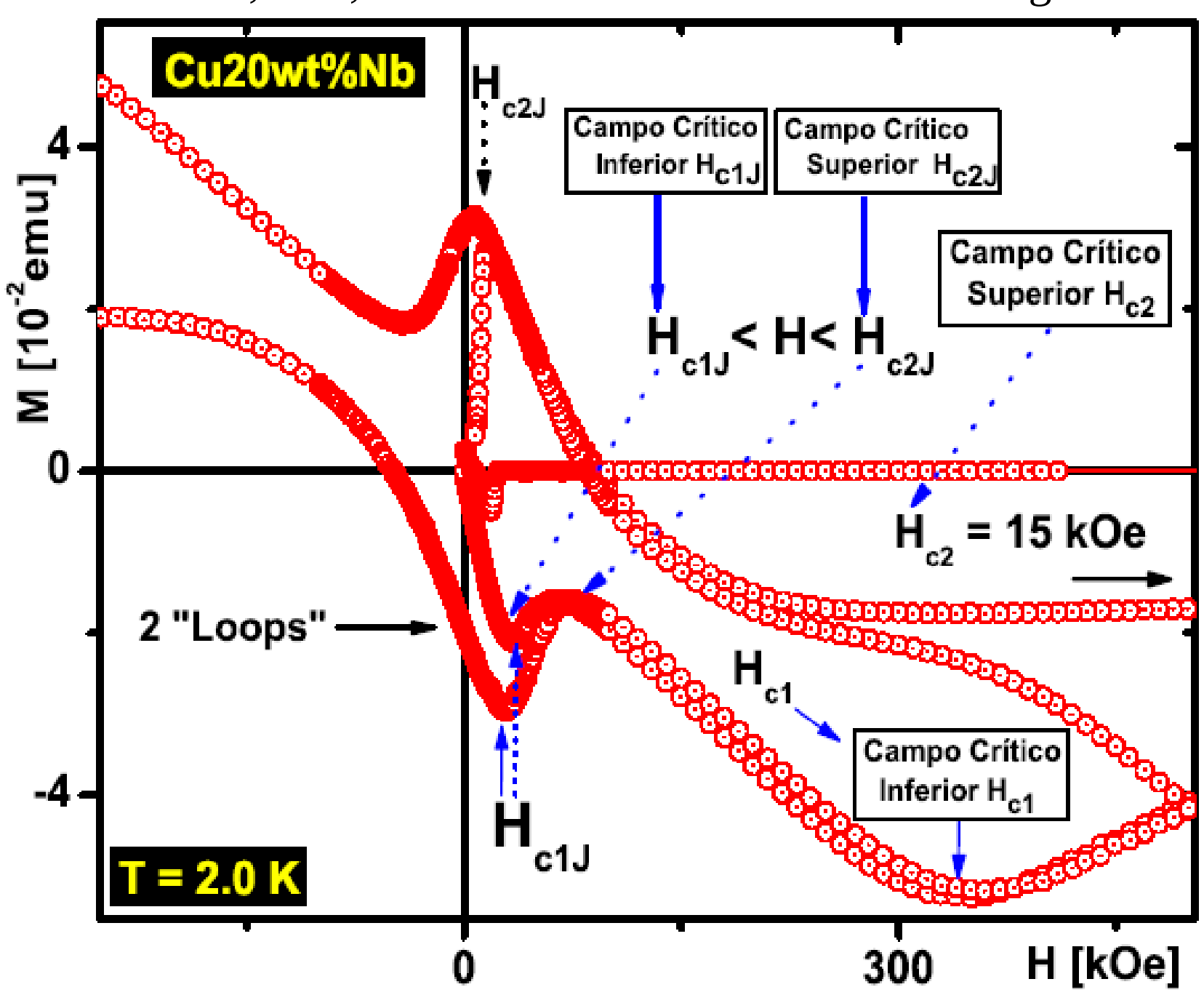

Figure 9: Two magnetization loops vs magnetic fieldmeasured in the sample Cu20wt%Nb. From this measure, it is possible to identify two regions for the grains of Nb. The behavior dominated by the Josephson coupling, whose magnetic response is in the region bounded by $H_{C1J} \leq H \leq H_{C2J}$ it is exclusively granular and the other region of volumetric response behavior of grains, which is delimited by $H_{C1} \leq H \leq H_{C2}$. The region of fields situated in $H < H_{C1J}$ is a region completely Meissner.

Based on the consideration of the existence of a LTSC coherence length $\xi_{LTSC}$ (whose value is large compared to interatomic distances) known, we can, through our experimental data, estimate the value of the coherence length $\xi^{Cuxwt\%Nb}$ for the samples Cuxwt%Nb. We will assume that in null temperature the coherence length ξt is given by [32]:

$$\xi = \frac{h V_F}{\pi \Delta} \qquad (6)$$

where, $V_F$ is the renormalized velocity of quasi-particles [32]. In the weak coupling regime $2\Delta = 3.5\, k_B T_c$. Thus, the formulation in 6 is valid in the limit where the free electronic medium path is much larger than $l \gg \xi$ . For simple metals (Nb), $V_F \approx 1 \times 10^8\, cm/s$ [36]. Thus, from the relationship 6 we can extract:

$$\xi^{Cuxwt\%Nb} \times T_C^{Cuxwt\%Nb} \approx 1\, \mu m \times K \qquad (7)$$

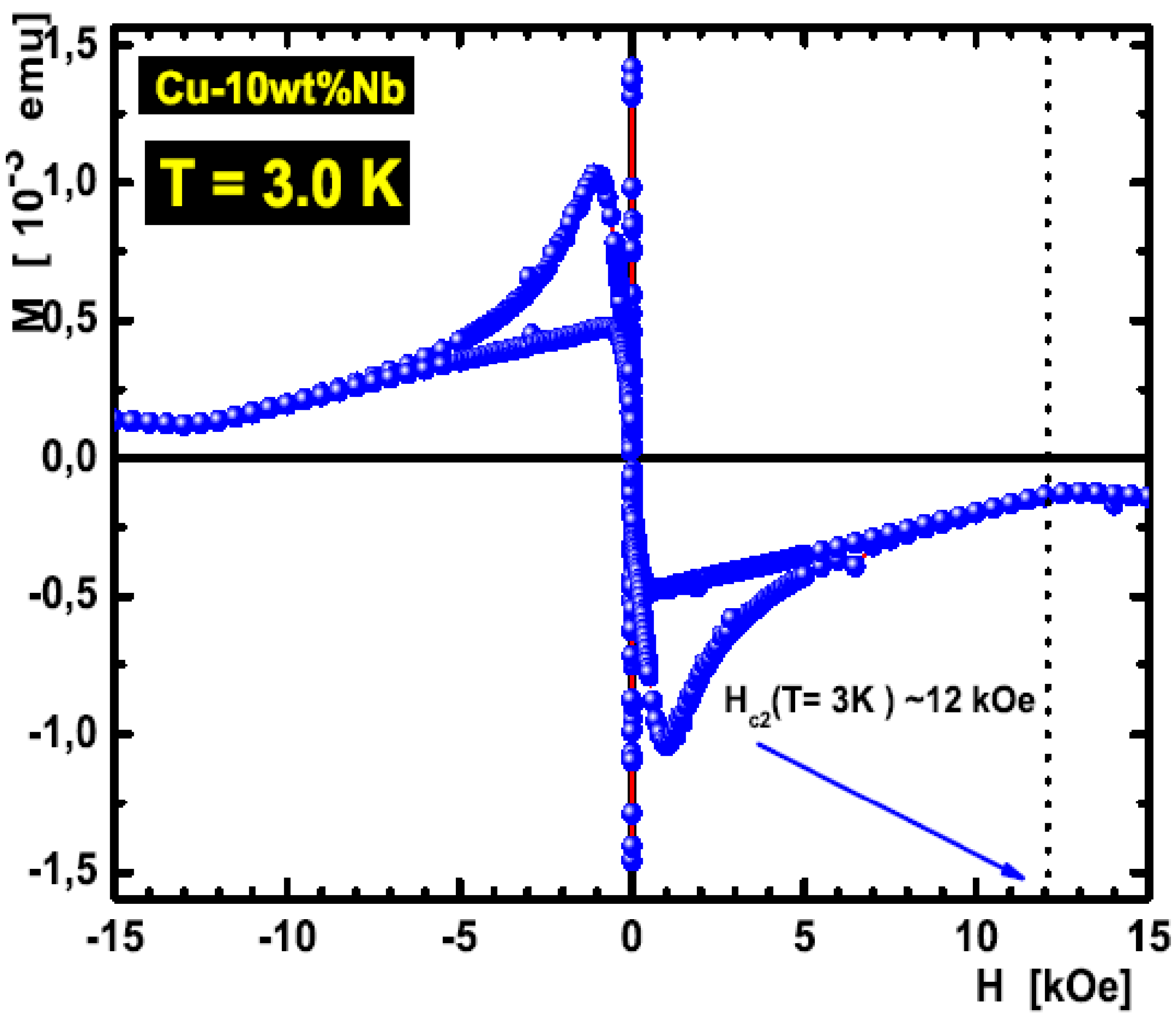

Figure 10: $M^{dc}[H,T,\theta]$ magnetization loop measured in the sample Cu10wt%Nb. The measure was taken under a constant temperature $T = 3\,K$. We can clearly see a completely (see inset at figure 11)Meissner region bounded by $H_{C1J} \leq H \leq H_{C2J}$. The volumetric response behavior of grains, which is delimited by $H_{C1} \leq H \leq H_{C2}$.

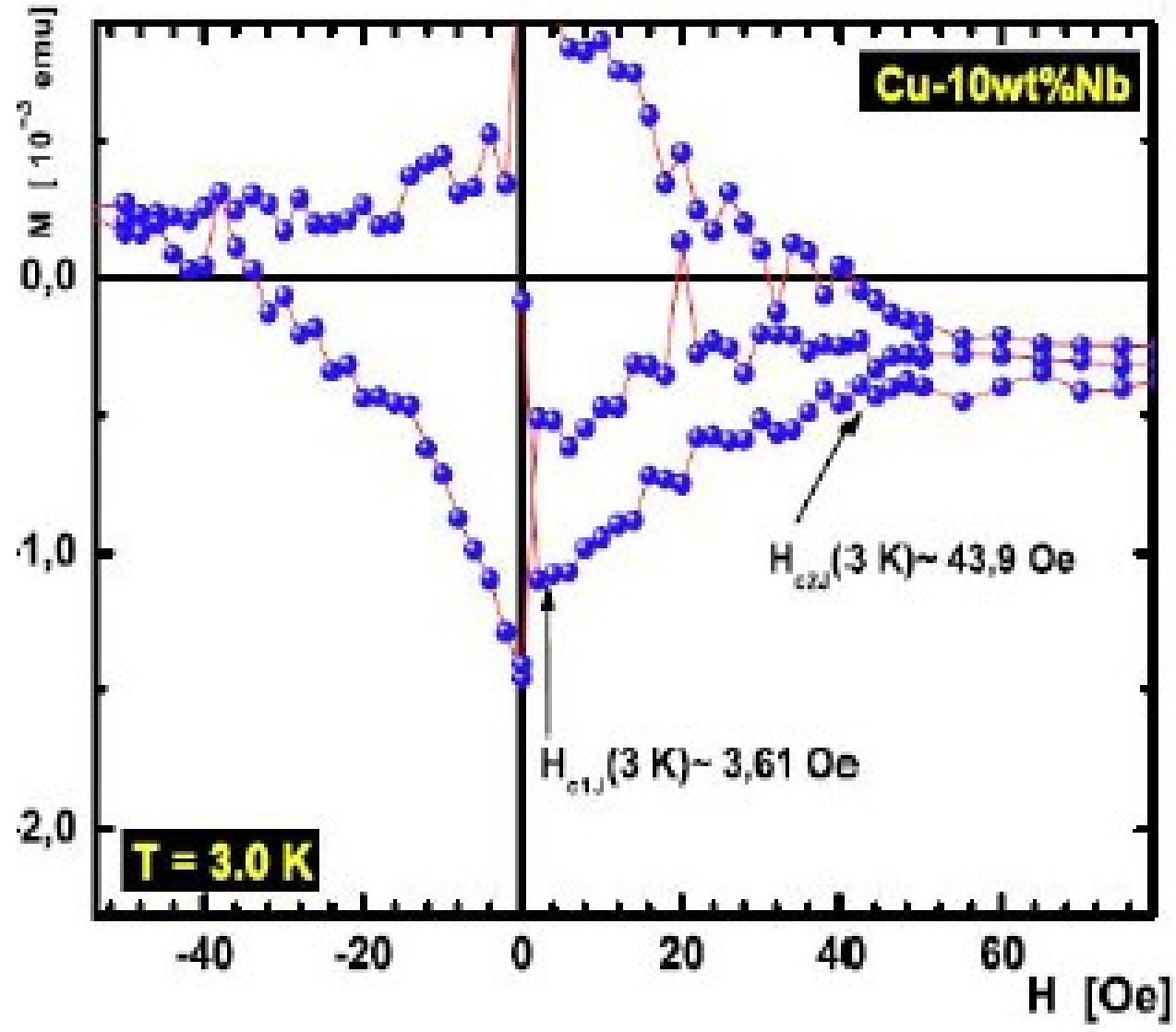

Figure 11: Inset of $M^{dc}[H,T,\theta]$ magnetization loop measurement of the figure 10 for the sample Cu10wt%Nb. The constant temperature was $T=3\,K$. We can identify $H_{C1J}(T=3\,K)=3.61\,Oe$ and $H_{C2J}(T=3\,K)=43.9\,Oe$.

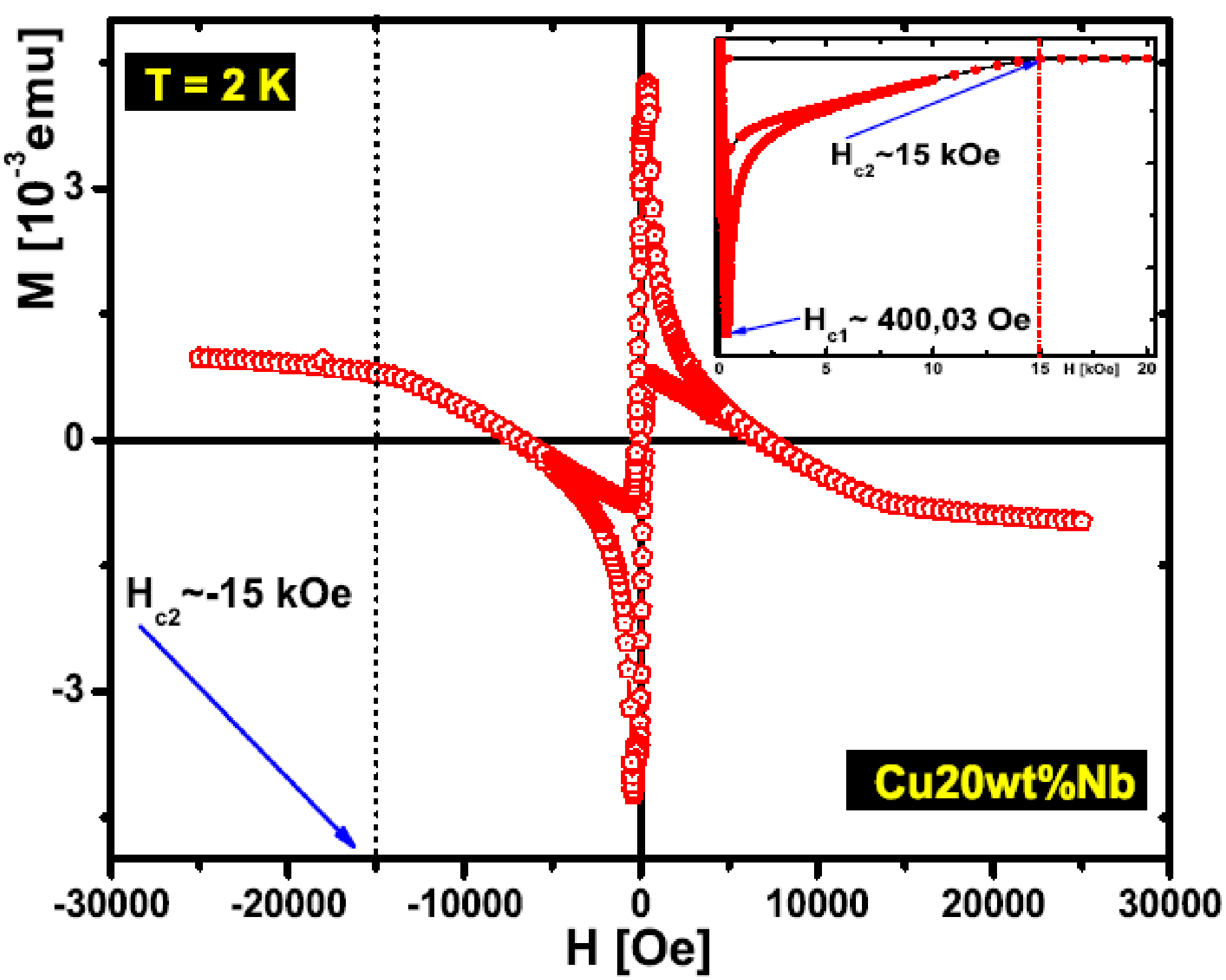

Figure 12: $M^{dc}[H,T,\theta]$ magnetization loop measured in the sample Cu10wt%Nb in magneticfields $-25000\,Oe \le H \le 2500\,Oe$.The constant temperature was $T=2\,K$.We can identify $H_{C1J}(T=2\,K)=99.93\,Oe$ and $H_{C2J}(T=2\,K)=199.25\,Oe\,(\text{see for more details the figure } 19)$. Also, we measured $H_{C1}(T=2\,K)=400.03\,Oe$ and $H_{C2}(T=2\,K)=15\,kOe$.

The equation (7) relationship is obeyed very well by metals simple [36] and predicts a well-known fact that$\xi$its inverselyproportional to$T_C$. Thus, we can estimate$\xi^{Cuxwt\%Nb}$using the transition temperature measured in out magnetotransportexperiments,$T_C = 8.2\,K$, replacing this value in 7, we will have:

$$\xi^{Cuxwt\%Nb} \times 8.2\,K \approx 1\,\mu m \rightarrow \xi^{Cuxwt\%Nb} \approx 0.122\,\mu m = 122\,nm (8)$$

Thus, from the calculation made in 8 we come to a value in$\xi \approx 1220\,\AA$, which is a typical value for LTSC's.Through equation 7 we can also estimate the correlation length $\xi$(T)for the temperature$T_{CJ}^{MXT}$of samplesCuxwt%Nb. To do so, we will use the temperature$T_{CJ} = 7.01\,K$ determined by the measurement$M^{dc}$xT of thegraph of the figure 1. This measure was carried out in a constantapplied magnetic field$M^{appl}$=15 Oe. Using equation 7,we will have

$$\xi^{Cuxwt\%Nb}(T_{CJ}) \times T_C^{Cuxwt\%Nb} \approx 1\,\mu m \rightarrow \xi^{Cuxwt\%Nb}(T_{CJ}) \times 7.01\,K \approx 1\,\mu m (9)$$
$$\rightarrow \xi^{Cuxwt\%Nb}(T_{CJ}) \approx 0.143\,\mu m = 143\,nm (10)$$

thus, we found the value $\xi^{Cuxwt\%Nb}(T_{CJ}) \approx 1430\,\AA$, for the granular coupling temperature, which It is 21 Ågreater than the coherence length from bulk pure Nb to samples Cu10wt%Nb-10-01 and Cu20wt%Nb-11-02.

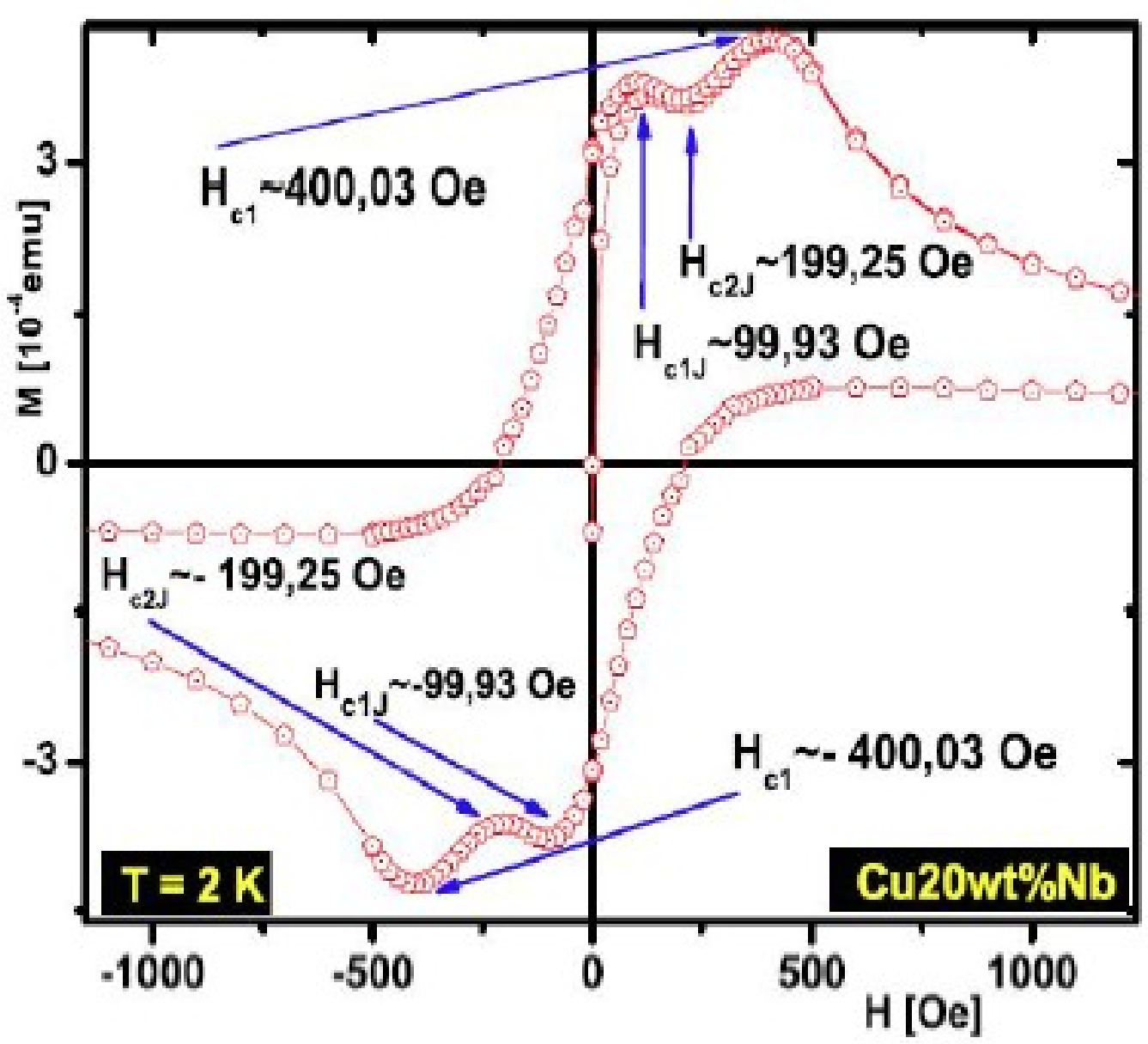

Figure 13: $M^{dc}$[H,T,$\theta$] magnetization inset from the figure 12 at $T = 2\,K$.We can identify $H_{C1J}(T = 2\,K) = 99.93\,Oe$, $H_{C2J}(T = 2\,K) = 199.25\,Oe \wedge H_{C1}(T = 2\,K) = 400.03\,Oe$.

It ispossible to determine the length of coherence$\xi$(T)through exclusively the superconducting parameters obtainedthrough the results of our magnetization measurements. If we turn our attention to the magnetic phase diagram shown in the figure 12, we can see that in the measure of $M^{dc}$[H,T,$\theta$ performed in $T = 2\,K$ can extracted the necessaryinformation for a good estimation of the coherence length of the superconducting order parameter$\Psi$through these valuesverified experimentally. For, as discussed, the coherence length $\xi_{LTSC}$of the LTSC has a relative large value when comparedto the interatomic distance of the material. To estimate the value of $\xi$(T) we will use the magnetizationmeasurement from the Cux20%Nb (sample-11-02) and we will use equation(A. A. Abrikosov [37]and R. D. Parks [36] ):

$$H_{C1J}(T)=\frac{\Phi_0}{2\pi\xi^2}\left(1-\frac{T}{T_c}\right) \quad (11)$$

which can be rewritten like this,

$$\xi^2(\mathrm{T})=\frac{\Phi_0}{2\pi H_{c2}(T)}\left(1-\frac{T}{T_c}\right) \quad (12)$$

For the sample Cux20%Nb-11-02 we have the following values of $H_{C1J}$,$H_{C2J}$,$H_{C1}$,$H_{C2}$and $T_C$:

- $H_{C1J}(T=2K)=99.93\,Oe$
- $H_{C2J}(T=2K)=199.25\,Oe$
- $H_{C1}(T=2K)=400.03\,Oe$
- $H_{C2}(T=2K)=15\,kOe$
- $T_C=8.2\,K$

Thus, through equation 12 we have the following values for $\xi(T)=2K$

1. ⇒in the region $H_{C2}(T=2K)=15\,kOe$:
   $\xi^2$(T)= (13)

Thus, in the region $H_{C2}(T=2K)$ we have $\xi^2$(T)=187.5Å

2. ⇒Determining $\xi_J(T=2K)$ in the region next to $H_{C2J}(T=2K)$ where $H_{C1J}(T=2K)<H_{C2J}(T=2K)<H_{C1}(T=2K)<H_{C2}(T=2K)$. Using equation 12 we will have
   $\xi_J(T=2K)=$ (14)

Thus, in the region $H_{C2}(T=2K)$ we have $\xi_J(T=2K)=1248\,Å$.

These results were already expected, as close to $H_{C2}(T=2K)$ the vortices lattice gradually loses its organization the long-range triangular arrangement and favors the organization of quadrangular of this Abrikosov lattice [37], whose lattice parameter varies with $B^{-1/2}$. And, if the lattice parameter increases, the phasecoherence length of $\Psi$decreases.We can also use the equations of F. London and H. London [28] to examine what happens to the depth of magnetic fieldpenetration $\lambda_0$in the upper critical field $H_{C2}(T=2K)$. The physical observable that defines depth magnetic fieldpenetration into the superconductor can be written as

$$\lambda=\frac{mc^2}{4\pi n_s e^2} \quad (15)$$

empirically, this equation can be rewritten as

$$\lambda(T)=\frac{\lambda_0}{\left[1-\left(\frac{T}{T_c}\right)^4\right]^{1/2}} \qquad (16)$$

We can get the value of $\lambda_0$ of two ways, the first one is directly measuring this observable.

| Magnetic Field Penetration Depth $\lambda_0$ for Some Superconductors | | | | | | | | | |
|---|---|---|---|---|---|---|---|---|---|
| Element (Metallic or HTSC) | Al | Cd | Hg | In | Nb | Pb | Sn | Tl | $YBa_2Cu_3O_{7-\delta}$ |
| $\lambda(0)$ in (Å) | 500 | 1300 | 380-450 | 640 | 470 | 390 | 510 | 920 | 1700 |

Table 2: Table containing information about the magnetic fieldpenetration depth $\lambda_0$for some superconductors. From the references E. A. Lynton, Superconductivity - London- (1962) [38] and Review of Modern Physics, G. Blatter *et al.,* (1994)[39].

Thesecond one is making an estimate through equation 15: If suppose that in$T=0$ all electrons are superconductors,this is It is,$n_s=n_0\approx 10^{22}$ and we use $m=m_e\approx 10^{-27}\,g$,$c\approx 3\times 10^{10}\,cm\times s^{-1}$, $e=1.6022\times 10^{-19}\,C$, weobtain from the equation 15 $\lambda_0=600\,Å$. However, a look at the table 2 it is sufficient to verify that the valuefor$\lambda_0$of the Nb presented is appreciably different of $\lambda_0=600\,Å$ . So wewill use the mean of these two values: $\lambda_0=535\,Å$.

Thus, by replacing the$\lambda_0$ in the equation 16, we will have

$$\lambda(T=2\,K)=\frac{\lambda_0}{\left[1-\left(\frac{T=2\,K}{T_c=8.2\,K}\right)^4\right]^{1/2}}=\frac{53{,}594\,nm}{\left[1-\left(\frac{2\,K}{8.2\,K}\right)^4\right]^{1/2}}\approx 53.594\,nm \qquad (17)$$

Thus, we estimate for the sample Cux20%Nb-11-02 at $T=2\,K$ T= 2 K, the values $\xi(T=2\,K)=187.5\,Å$and $\lambda(T=2\,K)=535.94\,Å$, we can then find the value of $\kappa=\frac{\lambda}{\xi}$for this sample through the relationship [37]:

$$\kappa=\frac{\lambda}{\xi}\Longrightarrow for\,T=2\,K\Longrightarrow\kappa(T=2\,K)=\frac{\lambda(T=2\,K)}{\xi(T=2\,K)}=\frac{535.94\,Å}{187.5\,Å}\approx 2.858 \qquad (18)$$

Observing the value of the result of the calculation performed in equation 18 and comparing with the results for the pure Nb [38] and [36], we will realize that there are very small differences in the values of$\lambda(T)\,\xi(T)$and $\kappa(T)$, so it's worth remembering here, phase diagram shown in the sample preparation process Cuxwt%Nb[14],[15][24–26], which gives us the information on the solubility of Nb in Cu$\rightarrow S_{Nb\rightarrow Cu}$ is around $1\,wt\,\%$at a temperature below $T_{1\,wt\,\%}^{(S_{Nb\rightarrow Cu})}$=1100 $^o$C, also we must to reinforce that this solubility decreases rapidly with the temperature. At room temperature this same solubility $S_{Nb\rightarrow Cu}$drops to less than $0.1\,wt\,\%$ , $T_{0.1\,wt\,\%}^{(S_{Nb\rightarrow Cu})}$= 300 K.In [14], the phasediagram indicates that the solubility of Cu in Nb $\rightarrow S_{Cu\rightarrow Nb}$at temperatures close to $T=1100^o$Cis around of $S_{Cu\rightarrow Nb}=¿$0.07 wt%. One of most important information which we have from our sample preparation method, is the consequence of this solubility/miscibility in the value change of $T_{c-Bulk}^{(Nb)}$and $T_{c-Bulk}^{(Cuxwt\,\%Nb)}$. $\Delta T_c=T_{c-Bulk}^{(Nb)}-T_{c-Bulk}^{(Cuxwt\,\%Nb)}=9.25\,K-8.20\,K=1.05$.

## II.III. MAGNETO-TRANSPORT MEASURES AT THE SYSTEM Cuxwt%Nb:$\rho^{dc}$[H,T,$\theta$], R$^{dc}$[V,I H,T,$\theta$], I[V,H,T,$\theta$] and $J_c$[H,T,I,$\theta$]

We performed resistivity and resistance measurements with angular dependence with the aim of verifying some anisotropic property of the Cuxwt%Nb granular system. However, before the start of all the transport measures, we have to couple to the circuit table where was the sample. The actual sample space zero magnetic fieldvalue does not always have null value as shown on the drive reader PPMS®electronic control system measurement software and displayed on the microcomputer screen. So, we identify the real residual magnetic field $H_{appl}^{residual}$at the superconducting PPMS®magnet (those experimental calibration data of the magnet superconductor residual field can be found elsewhere or at the supplementary material section for this manuscript). We did that to times. For these, two calibrations ($N=1,2$) of magnet magnetic residualfield, we found the following real values of internal magnetic field measured by Hall sensor:(a)$H_1^{residual}=0.19\,Oe$ Oe and at (b) $H_1^{residual}=0.23\,Oe$Oe. The average value of $H_{appl}^{residual}$is:$\frac{\Sigma_{i=1}^{N} H_i^{residual}}{N}=H_{appl}^{residual}=\frac{H_1^{residual}+H_2^{residual}}{2}=\frac{0.19\,Oe+0.23\,Oe}{2}=\frac{0.42\,Oe}{2}=0.21\,Oe$. Thus, aPPMS®magnetic field value correction in the superconducting magnet can be made by subtracting the actual measured residual magnetic field value$H_{appl}^{residual}=0.21\,Oe$ to reach $H_{0-Effective}^{appl}=H_{0-Nominal}^{appl}=0$. In another way,if the value was negative, the compensation to be made would be the addition of the measured residual magnetic field value with the Hall sensor. The curves of the figures 14 and 15 show several resistivity curves with angular dependence on the applied magnetic field and the larger plane of the sample, which has geometric shape of a strip, see information about geometric properties in the table 1). As can be seen in these curves, the resistance curves behavior is the same for all orientations of the sample surface in relation to the magnetic applied field. It is interesting to note that we do not do these measured with the interest of visualizing the double transition shown in the curves 7, 8, 9, 10, 11 and 12, therefore, in these resistance curves, which were carried out in a completely ac measurement process. We refine the parameters of the excitation ac field$h_{Exct}^{ac}$and it's oscillation frequency$\upsilon$, and the care and control of these parameters are decisive for avoiding any frustration of the phase of the superconducting order parameter of any grain need to remain non-random, to interfere constructivelywith the phase of any other grain. At the figure 14-b, we can display the value of the volumetric transition temperature of the ranged from $H_{appl}^{dc}$= 0 Tto $H_{appl}^{dc}$= 1 T a) Measure taken in $\theta=0^o$b) It is possible to observe the temperature value of volumetric transition of Nb grains,$T_{c-onset}^{(Cux\,20\,\%Nb)}=8.2\,K$.

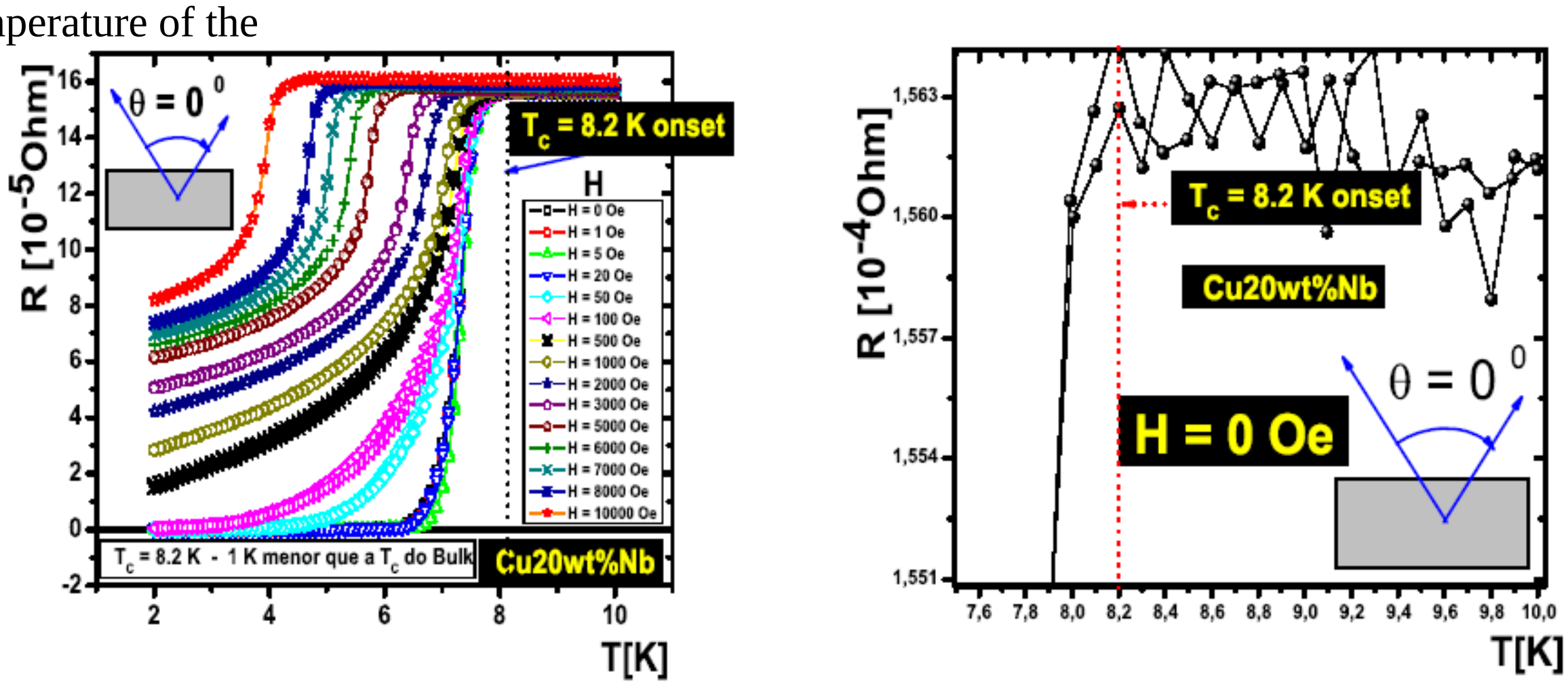

Figure 14: Measures electrical resistance to the sample Cu20wt%Nb performed through of the ac mode without strict control of the parameters $h_{Exct}^{ac}$and its oscillation frequency $\upsilon$. The magnetic applied fields in these measures

grains of Nb$T_{c-onset}^{(Cux\,20\,\%Nb)}=8.2\,K$, as discussed, this value different from the volumetric transition temperature of Nb($T_{c-Bulk}^{(Cu\,20\,wt\,\%Nb)}=9.25\,K$) pure because there is, in the case of samples Cuxwt%Nb an amount of Cu dissolved in Nb (also possible Nb dissolved in Cu). The most important new information which we can draw from the measures (confirmedby the structural analysis by scanning electron microscopy) of the figures 14 and 15 is that there is no evidenceof granular anisotropy in the magneto transport$R^{dc}(V, H, T, \theta)$properties of the sample Cu20wt%Nb. If we look at the figures 14 and 15of electrical resistance measurements, we will obtain approximately to $R^{Cu\,20\,wt\,\%Nb}\approx 1.56\times 10^{-4}\,\Omega$for the highest resistance saturation value for $H_{appl}^{dc}$= 0 and $\theta=0^o$(note that thesevalues are normalized). For the superconductorsmaterials, if the resistivityis not exactly zero, in the presence of small currents$J\longrightarrow 0$, this will have values close to $\rho^{Cu\,20\,wt\,\%Nb}\approx 10^{-24}\,\Omega cm$seefor example, information on these values ??in the references [36] [40]). This way we can introduce into theresults of the 20 e 21 the geometric parameters of our samples to try to estimate the non-standard value of theresistance. The geometric properties of the sample Cu20wt%Nb-11-02 what are, $l\times w\times t=0.40\,cm\times 0.20\,cm\times 0.01\,cm$(see table 1 for the sample Cu20wt%Nb -11-02), we can get the resistivity values:

$$R=\frac{\rho\times L}{A} \qquad (19)$$

where $L$ is the length of the sample and $A$ is the section area cross-section through the electric current, thus to $T\leq T_c$,

$$R_{Cu\,20\,wt\,\%\,Nb}=\frac{1\times 10^{-24}\,\Omega cm\times 0.20\,cm}{0.01\,cm\times 0.40\,cm}\approx 5\times 10^{-24}\,\Omega \qquad (20)$$

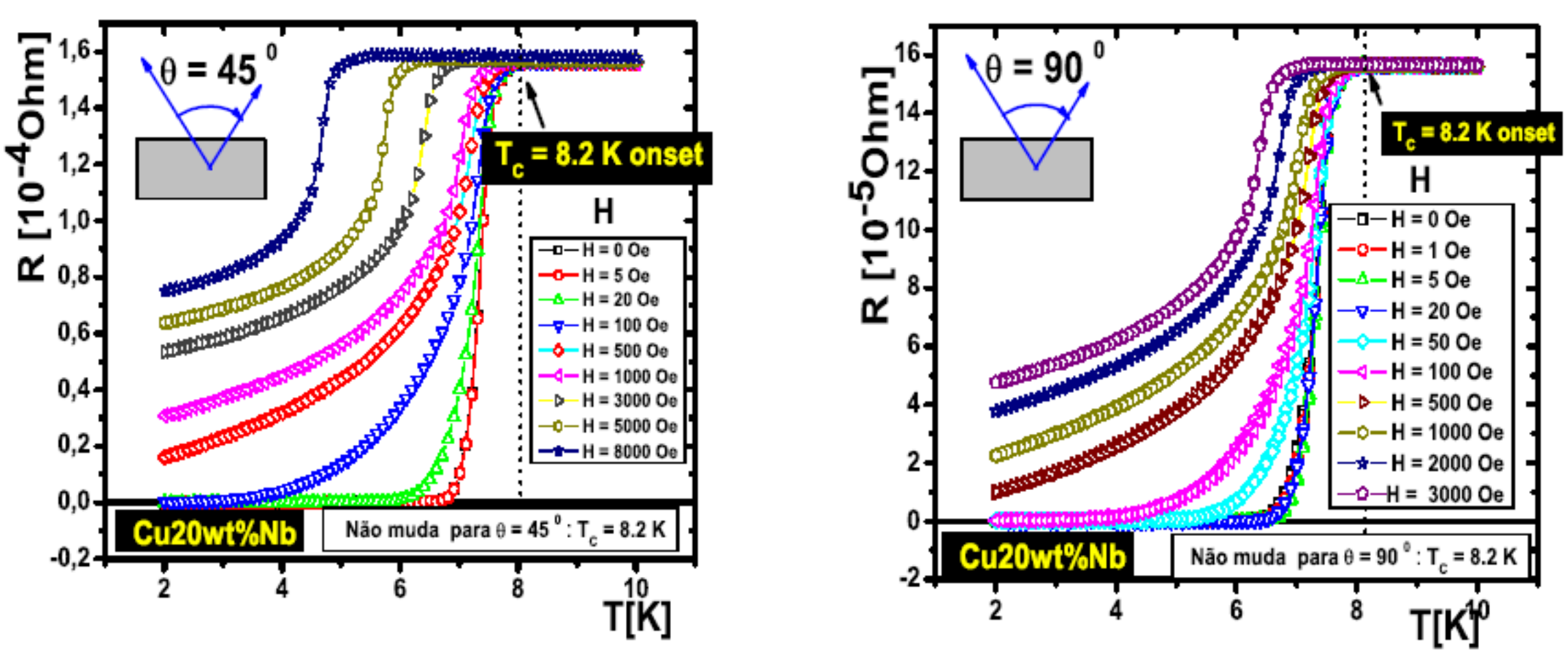

Figure 15: Measures electrical resistance $R^{Cu\,20\,wt\,\%Nb}(H, T, h_{Exct}^{ac}, \upsilon, \theta)$to the sample Cu20wt%Nb performed through of ac mode with the control of the shaking transverse magnetic field $h_{Exct}^{ac}$and its oscillation frequency$\upsilon$.

The external dc applied magnetic field ranged from $H^{dc}_{appl}$= 0 Tup to$H^{dc}_{appl}$= 1 T : a) Measuretaken in$\theta=45^{o}$. b) Measure taken in $\theta=90^{o}$.$T^{(Cux\,20\%Nb)}_{c-onset}(0)=8.2\,K$.

From these values, we will verify the proposed model by Maxwell in Treatise on Electricity and Magnetism, James Clerk Maxwell (1873) [41] where is it possible to estimate the resistivity value of alloys composed of different resistivity values, and then we will compare with the results for the sample system CuXwt%Nb (particularly, the result obtained in equation 20 to the sample Cu20wt%Nb-11-02 exactly at the transition point, ($T^{(Cux\,20\%Nb)}_{c-onset}(0)=8.2\,K$). The Maxwel's expression for the resistivity of an alloy containing two different metallic materials with different properties is [41]:

$$\rho_{alloy}=\frac{2k_1+k_2+p(k_1-k_2)}{2k_1+k_2-2p(k_1-k_2)}\times k_2 \qquad (21)$$

where $\rho_{alloy}$is the resistivity of the alloy, $k_1$ is the resistivity of material 1, in quantity smaller, $k_2$ is the resistivity of material 2, in greater quantity, and p is the fraction of the constituent $k_2$ in the alloy. Thus, bydoing the following replacements:$\rho_{alloy}$by $\rho^{alloy}_{(Cux\,20\%Nb)}$, $k_1$ by $\rho_1$, $k_2$ by $\rho_2$and$p$ by $f$ which is the fraction of Nb by weight (xwt%=20wt%) at the sample Cu20wt%Nb,  we will have the following equation:

$$\rho^{alloy}_{(Cux\,20\%Nb)}=\frac{2\rho_1+\rho_2+f(\rho_1-\rho_2)}{2\rho_1+\rho_2-2f(\rho_1-\rho_2)}\times\rho_2 \qquad (22)$$

Now, replacing $\rho_1$by$\rho_{Nb}$and $\rho_2$by$\rho_{Cu}$at the equation 22, we will have:

$$\rho^{alloy}_{(Cux\,20\%Nb)}=\frac{2\rho_{Nb}+\rho_{Cu}+f(\rho_{Nb}-\rho_{Cu})}{2\rho_{Nb}+\rho_{Cu}-2f(\rho_{Nb}-\rho_{Cu})}\times\rho_{Cu} \qquad (23)$$

With some algebra, we come to:

$$\rho^{alloy}_{(Cux\,20\%Nb)}=\rho_{Cu}\times \qquad (24)$$

Now considering that at temperatures $T\approx T_c$,$\rho_{Nb}=0$, the equation 24 is simplified to

$$\rho^{alloy}_{(Cux\,20\%Nb)}=\rho_{Cu}(1-3f)\ \text{ for } T\approx T_c \qquad (25)$$

We will use the results of the $\rho_{Cu}$of the reference [42], where was measured$\rho_{Cu}\approx 2.12\times10^{-9}\,\Omega\,cm$ :

$$\rho^{alloy}_{(Cux\,20\%Nb)}=\rho_{Cu}(1-3f)=(2.12\times10^{-9}\,\Omega\,cm)\times[1-(3\times0.2)]=(2.12\times10^{-9})\times0.4=8.48\times10^{-10}\,\Omega\,cm$$

for $T\approx T_c$(26)

Comparing the two results, the estimated by [36] and [40] for $\rho\approx1.0\times10^{-24}\,\Omega\,cm$, and the value calculated inMaxwell's equation 25 nearthe transition point ($T^{(Cux\,20\%Nb)}_{c-onset}(0)=8.2\,K$is $\rho^{alloy}_{(Cux\,20\%Nb)}\approx8.48\times10^{-10}\,\Omega\,cm$), we immediatelyperceive a difference of values.

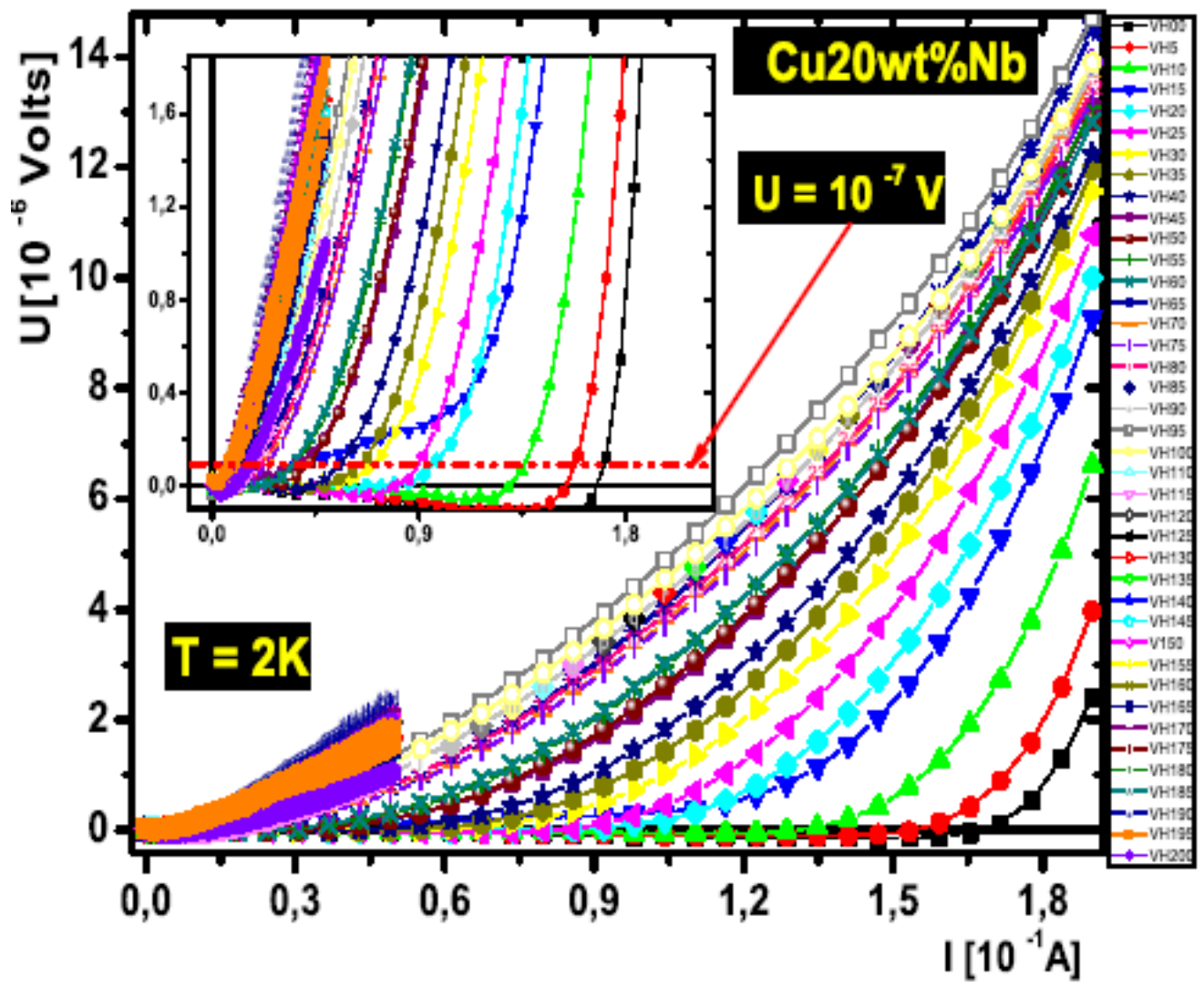

Figure 16: Measures of current vs voltage, $I \times V$ curves, in various different $H_{appl}^{dc}$ applied magnetic fields under a constant temperature $T=2\,K$.

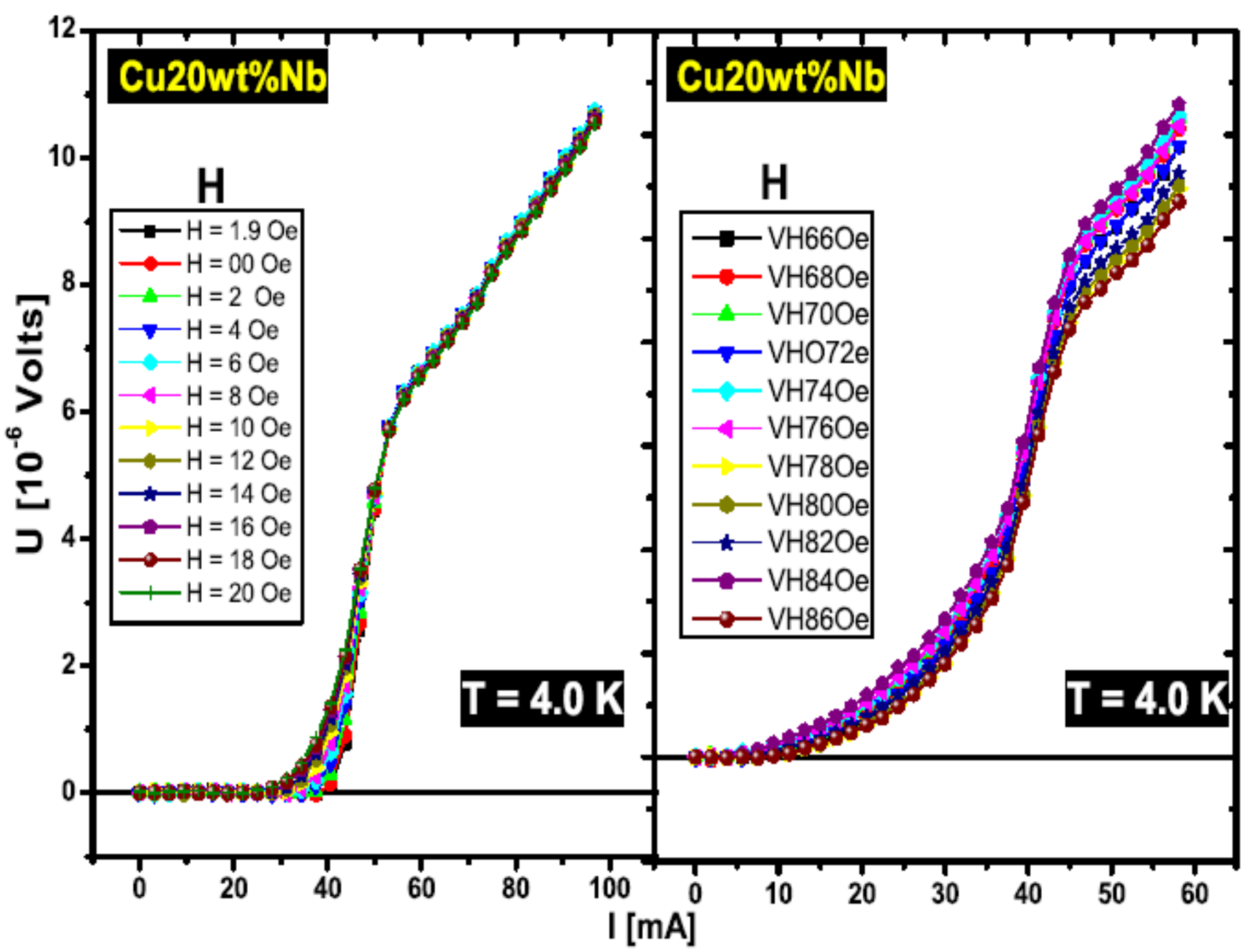

Figure 17: Electrical current vs voltage, $I \times V$ under a constant temperature $T=4\,K$. We used various different external applied magnetic fields $H_{appl}^{dc}$ intensities. These measurements, $I \times V$ were very important so that we could extract the spectrum of the $I_c^{Cu\,20\,wt\,\%Nb}$, points, with a cut-off voltagecriterion: $V_{cut-off}^{Cu\,20\,wt\,\%Nb}=10^{-7}\,V$.

The explanation for this is that in normal metals such as Cu and other metals, the dependence of the resistance (resistivity) with temperature, for temperature values considered low, is cubic as suggested by Debbye's theory [43], and not linear as suggested by Maxwell to obtain the resistivity of alloys [41] without distinction of temperature regimes.

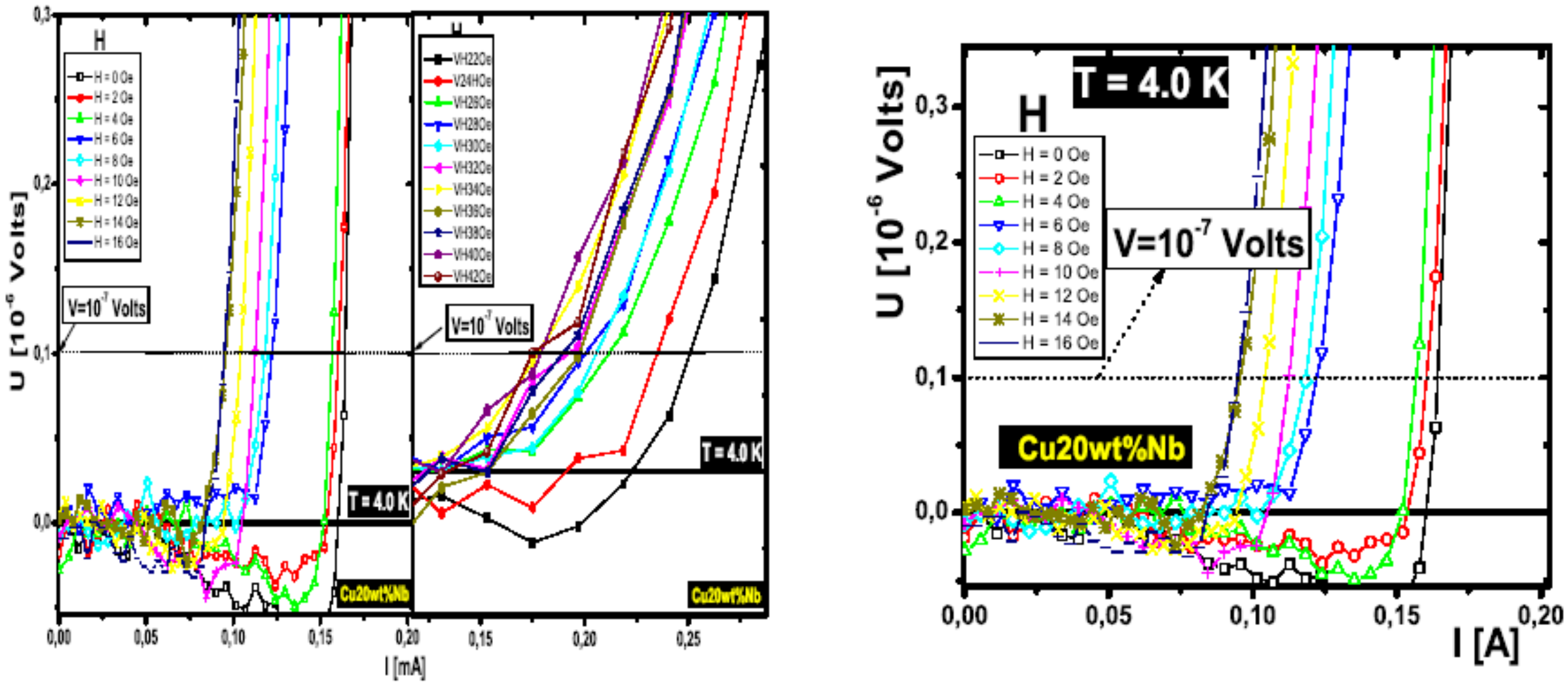

Figure 18: $I \times V$ measurements. $T = 4\,K$. The cut-off voltage to extract the spectrum of the $I_c^{Cu\,20\,wt\,\%Nb}$ points is: $V_{cut-off}^{Cu\,20\,wt\,\%Nb} = 10^{-7}\,V$.

At the measures at figures 17 and 18 we make an estimation for the value of the critical current in the

Cuxwt%Nbsamples. The process of obtaining these measures was completely ac and dc. As we know, superconducting materials do not really presented losses in the transport properties only under dc direct current conditions. Under alternating current conditions superconductors dissipate power through magnetization and demagnetization of the material, in an analogous process to hysteresis. These losses are, for example, much smaller than the resistive losses on copper conductors of the same capacity. Even considering such dissipative losses, we used the data on the figures 17 and 18 tofind the value of the critical current density for the two samples Cu20wt%Nb and Cu10wt%Nb. The values obtained were (a) $J_C^{Cu\,20\,wt\,\%Nb}(H) \approx 406\,A/cm^2\,at\,T = 2\,K$ and (b) $J_C^{Cu\,10\,wt\,\%Nb}(H) \approx 152.5\,A/cm^2\,at\,T = 4\,K$. In magneto transport measurements, we use a device of magnetic fieldmeasurement, consisting of a sensor with 10 Hall bridges, that is, a conductor with 10 meters of transverse electrical potential difference $V_{transverse}^{Hall}$. The device driver has straight section of dimensions $5.0 \times 0.2 \times 0.01\,mm^3$, with the 10 pairs of contacts regularly distributed over the length of the sensor. This driver was put on the PPMS®ACT Puck circuit together with the sample and connected in series therewith. Once the circuit was installed, we inserted it inside the cryostat to the initiation of the measures and began the process of calibrating magnetic fieldvalues. After this calibration, we began the process of measurements of the curves $I \times V$, as shown in the figures 16, 17 and 18. These current vs voltage spectra data measurements, $I \times V$ curves, were very important so that we could extract the points ($I_C^{Cu\,10\,wt\,\%Nb}(H)$), with a cut-off voltage criterion of $V_{cut-off}^{Cu\,20\,wt\,\%Nb} = 10^{-7}\,V$, as can be seen more clearly in figure 18. The Hall sensor allowed us to use the real effective applied magnetic fieldvalues $H_{appl}^{dc}$ to take our data curves $I \times V$ and consequently, to investigate more quantitative aspects of behavior of the critical current $I_C^{Cu\,10\,wt\,\%Nb}(H)$ as a function as a function of $H_{appl}^{dc}$. From this construction, and of the criterion of measures adopted(with $V_{cut-off}^{Cu\,20\,wt\,\%Nb} = 10^{-7}$), next we willshow among other results, oscillations in the critical current measured as a function of the magnetic appliedfield, which is unprecedented result for this type of arrangement three-dimensional superconducting grains dispersed in normal matrix.

## III. DISCUSSION: QUANTUM INTERFERENCE AND COHERENCE EFFECTS IN THE SUPERCONDUCTOR SYSTEM Cuxwt%Nb

As it was seen, in granular superconductors, the properties are below critical temperature of the bulk, $T_c^{bulk}$(H). However, there are many disorder effects [23] in the transition of coherence $T_{CJ}(H)$due to the possible randomness of the counts (Cu barriers in the samples Cuxwt%Nb). And such effects can be well described by a non-linear differential equation from an analogy to the Josephson phase difference in a single junction. This description allows us to handle the Josephson coupling effects under conditions of possible (and unknown) randomness of the points of contact and also allows us to understand the behavior of the criticalcurrent as a function of the magnetic field$H_{appl}^{dc}$ in two ways:1) Through a single peak in the current total criticism centered on $H_{appl}^{dc}=0.$2) Creating paths of between the junctions that may be oriented randomly, that it should provide a finite contribution to the electric current density at high magnetic applied fields. Under certain conditions, the normal part (Cu) may present a kind of Meissner effect in many low magnetic fields, with the increase of the magnetic field, the superconducting properties induced in the normal part of the junction by the superconducting material (Nb), in particular the Meissner effect are progressively destroyed. In practice, this destruction occurs inmagnetic fields that are significantly smaller than the critical fields $H_C$ *of the*superconducting part (Nb). For this reason, in addition to the Nb in Cu under certain temperature conditions during the preparation of the samples (Cu-Nb phase diagram), which we can say than in the case of a Josephson junction where one of the materials constituents is a normal metal, there are superconducting properties very weak on the normal side of the junction. The critical fields exclusively due to the superconductor material (Nb), in particular the larger critical field, have a tendency to be heavily modified due to the effects of proximity to the normal part (Cu). And this due the effects of proximity (we did not showdata where the external magnetic fieldis extremely high). According to [36], there are also a strong influence of the normal metal in the critical field$H_C$. These granular samplesCuxwt%Nb present as characteristics$a_{grain} \gg \lambda(T)$, where $a_{grain}$is the radius of the grain (as seen in images of scanning microscopy and the estimation of $\lambda(T)$. The characteristics of this type of granular arrangement where the grains are coupled by Josephson effect differs drastically from properties of granular films where the grains are normally much smaller and load effects (Coulomb repulsion) tend to destroy the Josephson coupling and this is a complication for the construction and analysis of such systems. But, on the other hand, our samples do not admit a continuous phenomenological description, where the superconducting phase can be designed to vary smoothly from one superconducting grain to the next. In fact, the Josephson interaction connects the different grains by the phases of their superconductingorderparameters is independent of the randomness of orientation of the contacts, since, of course, the real grains, Nb that we perform, are not “perfect spheres”, and this reality brings the randomness of the junctions in different pairs of grain along the Cu matrix. It is possible to model such granular systems assuming granular individual phase as$\phi_i$in the presence of anexternalfieldmagnetic field for grains small enough. We cannot consider that for grains of very large dimensions, or from the order of the parameters interatomic effects, load effects become very relevant, and the phase of the order parameter can become a variable completely dynamic, due to the Coulomb repulsion. The order parameter module$\Psi \vee$ can be assumed to be constant, with smooth variations only in the individual granular phases $\phi_i$ which then have dynamics degrees of freedom. Thus, the Hamiltonian [44] model that can describe this granular system is

$$H=\sum_{i,j>J_{ij}\cos(\varphi_i-\varphi_j-A_{ij})} \square \quad (27)$$

with

$$A_{ij}=\frac{2\pi}{\Phi_0}\int_i^j A.dl \quad (28)$$

representing the line integral of the magnetic vector potential $A$ between any neighboring grains $j$ at matrix, where

$$B=\nabla\times A \qquad (29)$$

$J$is the Josephson coupling energy between the neighborsgrains. In many cases in these systems, the shielding is weak and the fluctuations in the potential vector can be neglected. Some effects of disorder can be inserted into such systems by this model through random coupling between neighboring grains $J_{ij}$. However, in this model do not take account of changes in signals. In this case, these coupling constants$J_{ij}$are assumed to be the same as and equal to $J$. Another possibility of frustration in these systems is via randomness of the potential vector $A_{ij}$, by positional disorder of the grains in the matrix (irregular distribution of the grains in the matrix). For magnetic fields high enough, $A_{ij}$tends to vary over any interval between$[0,2\pi]$and this may in fact cause a signal change in the coupling between the neighboring grains producing a considerable amount of frustration in the coupling we can detect in our experiments. Nevertheless, we have adopted another way of modelling the granular system Cuxwt%Nb we produce.

We should emphasize in our systems is not is precisely the nominal Josephson coupling energy, but rather the gauge invariance of this considering the phases of the order on all the grains in the sample (not exactly monomodal). Therefore, we understand that order parameter$\Psi \vee$ or amplitude of condensation will either unite the set of spatial vectors, which in the real situation are randomly due to the fact that the junctions are not “ideal”. It is usual for these systems, assume that the Josephson coupling energy is [36,39]:

$$E_J=\frac{\Phi_0 I_J}{2\pi} \qquad (30)$$

where *IJ* is the critical current of the junction. And we can assume that in fact, through equation 30 the Nbgrains are coupled with the neighbors in the presence of magnetic fields and currents electrical connections. Ifthe system temperature complies with relationship $k_B T \gg E_J$, the vectors stillare will be randomly oriented. But when$k_B T \ll E_J$(where $k_B$is Boltzmann's constant), we can expect that the vectors in the regions of the junction contacts will effectively be oriented in a preferential direction. And this orientation will result in phase coherence over the entire granular volume of the sample. We, we can expect the behavior we saw in the data of the figures 1-6, where the coherent transition occurs at temperatures $T_{cJ}^{MXT}(H)<T_c^{MXT}(H)$(or $T_{cJ}^{\chi}(H)<T_c^{\chi}(H)$). For to estimate the coupling energy, we need just do $k_B T=E_J$. We can do this for the sample Cu20wt%Nb through of the $M\times T$measures figures 1 and 2 (and $T_{cJ}^{\chi}(H)\, the\, figures\, 9-12$). As value of the Boltzmann constant $k_B=R/N_A=0{,}0862\, Mev\, K=1.3807\times10^{-23} J/K$, we will have:

$$E_J\approx k_B T_{cJ}^{MXT,\chi}(H)(31)$$

in accordance with the data at the figures 1, 2 and $9-12$ we have the following situations:

1. $T_{cJ}^{MXT}(H) \Longrightarrow 7.31\,K \leq T_{cJ}^{MXT}(H) \leq 7.64\,K$: For dc magnetic fields $15\,Oe \leq H_{appl}^{dc} \leq 150\,Oe$ andusing equation 31 we will have:

for $T_{cJ}^{MXT}(H) = 7.31\,K$:

$$E_J^{MXT} \approx k_B \times T_{cJ}^{MXT}(H) \Longrightarrow E_J^{MXT} \approx 1.3807 \times 10^{-23}\,J/K \times 7.31\,K \approx 1.0093 \times 10^{-22}\,J (32)$$

for $T_{cJ}^{MXT}(H) = 7.64\,K$:

$$E_J^{MXT} \approx k_B \times T_{cJ}^{MXT}(H) \Longrightarrow E_J^{MXT} \approx 1.3807 \times 10^{-23}\,J/K \times 7.64\,K \approx 1.0549 \times 10^{-22}\,J (33)$$

2. $T_{cJ}^{\chi}(H) \Longrightarrow 6.36\,K \leq T_{cJ}^{\chi}(H) \leq 6.87\,K$: For the constant values$\upsilon = 75\,Hz$,$=h_{exc} = 5\,Oe$and magnetic fields with values $1\,Oe \leq H_{appl}^{dc} \leq 5\,Oe$

for $T_{cJ}^{\chi}(H) = 6.36\,K$:

$$E_J^{\chi} \approx k_B \times T_{cJ}^{\chi}(H) \Longrightarrow E_J^{\chi} \approx 1.3807 \times 10^{-23}\,J/K \times 6.36\,K \approx 0.87812 \times 10^{-22}\,J (34)$$

for $T_{cJ}^{\chi}(H) = 6.87\,K$:

$$E_J^{\chi} \approx k_B \times T_{cJ}^{\chi}(H) \Longrightarrow E_J^{\chi} \approx 1.3807 \times 10^{-23}\,J/K \times 6.87\,K \approx 0.94854 \times 10^{-22}\,J (35)$$

Thus, from the results calculated in equations 32, 33, 34 and 35 of items 1 and 2, we establish the following relations:

For measures for $T_{cJ}^{MXT}(H)$:

$$1.0093 \times 10^{-22}\,J \leq E_J^{MXT} \leq 1.0549 \times 10^{-22}\,J (36)$$

For measures for $T_{cJ}^{\chi}(H)$:

$$0.87812 \times 10^{-22}\,J \leq E_J^{\chi} \leq 0.94854 \times 10^{-22}\,J\, 0.94854\ (37)$$

To conduct more precise quantitative discussions on the results above $E_J$,we would need to allocate attention to the following aspects of measurement:$i$Carry out measurements, $M \times T$ and $\chi \times T$longer fieldranges. $ii$Decrease as much as needed the steps$\Delta H_{dc}$between a $M \times T$ and $\chi \times T$ measures, especially in the $M \times T$ experiments, where the variation of the fieldbetween two measurements was greater than 20 Oe.However, qualitatively, we can explain difference in values between $T_{cJ}^{MXT}(H)$, $E_J$and $T_{cJ}^{\chi}(H)$. We can to use the idea of critical state models to understand what in fact happens to the extent of susceptibility. This problem,models of susceptibility measurements, it has been treated in original form a few years ago [45,46], after the discovery of superconductors high temperature critical [17].

here

State Models are based on the very plausible hypothesis that each point of a sample subjected to a magnetic field, it is always the largest shield current possible, *e.g.*, the local critical current. This current, in turn, depends on the value of the field, which makes explicit the need to find an expression for $J_c(H)$.The Bean Model [47], the simplest of this class, assumes that the profile of the fieldin the sample is linear. Among others not so elementary possibilities, the Exponential State Model Critical (MEEC) [34]which was successfully used to adjust experimental results of sintered samples. If we assume that the fieldis small enough to ensure that the critical current of the inter granular material is not exceeded at low temperatures (in the case of susceptibility measurements $\chi^{ac}[T, H, h_{Exc}, \upsilon, \theta$in sample *Cux20%Nb*, we use a maxfieldof $H_{appl}^{dc} = 5\,Oe$), currents of intra-granular shielding prevents fieldentry into grains, while inter granular currents flow through the sample to ensure a zero magnetic fieldinside. The regions of temperatures immediately below$T_c(H)$,they will be restricted to the intra granular contribution. This situation persists and the resulting susceptibility re-enters $T_{cJ}^{MXT}(H)$, and $T_{cJ}^{\chi}(H)$,while remaining negative over the entire range of temperatures below $T_c(H)$. In this condition, $\chi$'reentering causing a reduction of $T_{cJ}(H)$. This reentrant effect is typical in *2D* networks of Josephson junctions. One way to properly model the order parameter for these transitions, in granular systems such as*Cux20%Nb* discussed above, was proposed by [23], which consists of assuming the superconducting order parameter as follows

$$\psi = \langle e^{i\phi} \rangle \quad (38)$$

where the brackets denote a kind of average over thermal energy and $\phi$ and the phase (which floats in real systems) of the superconducting phase of each grain individually. Maybe one problem that can be raised is that in the case of superconductors the translational order is not necessarily invariant. But in the case of our samples, due to the spherical geometry of the grains, the system is notprovided with structural translational anisotropy which may affect the properties of coherence superconducting system. We detected this with our magneto transport. $R^{dc}(V, H, T, \theta)$ measurements angular dependence,$\theta=0^o$, $\theta=45^o$ and $\theta=90^o$, as can be seen in the graphics of the figures 14 and 15 ($T_{c-onset}^{(Cux\,20\%Nb)}=8.2\,K$ *for those three angular orientations*). The nominal strength coupling$E_J^{M\times T}(E|J^{\chi})$from one junction to another, since we know that the grains do not exactly the same size and shape, but a distribution. We expect modulating the totaldensitycurrent modulated by the magnetic fieldis exactly the independent of the appliedfielddirection ($\theta=0^o$, $\theta=45^o$ and $\theta=90^o$) in the samples related its $\hat{c}=\hat{a}\times\hat{b}\,direction$. The junctions are in some way driven by the fieldwithout making any distinction about the particularity of each junction. Thus, we can use instead of the order parameter phase, a pseudo-phase of this, which develops a function well similar to the Josephson phase difference conventionally used (see[23]). From the many sources of granular disorder in our samples, no doubt the orientation of the junctions (Cu) deserves special attention. This is, in our analysis we assumed that $I_J$has the same value for all junctions. Thus, if$\vec{a}$is the vector joining two any neighborsNb grains, we can consider its module constant, but we cannot neglect the fact that $P(\theta)d\theta\sin(\theta)d\theta$to form an angle $\theta$and$\theta+d\theta$between the direction of two vectors in two any grain of Nb throughout the sample. Thus, when we apply the external magnetic field, we understand that a microscopic magnetic field$\vec{h}=(\vec{r})$ is created immediately in the regions around the grains, which may to some extent diminish the value of the superconductive correlation in these regions. If to consider

$$\vec{B}(\vec{r})=\bar{\vec{h}}(\vec{r})=\vec{\nabla}\times\vec{A} \quad (39)$$

where the bar ($\bar{\vec{h}}$indicates a lower effective volume considered, that is, an average taken over a small granular volume when compared to the volume of the entire sample *Cuxwt%Nb*.But, this volume considered to be large enough to contain a large number of grains. If we consider that there is no Josephson coupling, so

$$\vec{B}(\vec{r})=\mu\vec{H}\approx(1-f)\vec{H} \quad (40)$$

where the value$f$ can be seen as a factor of fill in the granular sample, that is, $f$ with value large in the small volume considered, means a larger collection of grains. Thus, in the presence of magnetic fields and currents different from zero, but of low values, let to] write free energy for unit ofvolume:

$$F=\overline{F}_0+\overline{F}_1+\frac{B^2}{8\pi\mu}(41)$$

where$\overline{F}_0$ is free energy in the absence of magnetic fields and currents. From $\overline{F}_1$ we can say that if the order parameter $\psi=\langle e^{i\phi}\rangle \longrightarrow \vee \psi \vee \langle e^{i\tilde{\phi}}\rangle$, where $\tilde{\phi}$is the new phase of order parameter that describes the correlations of the $\phi$. It is plausible to assume that it will contribute to free Josephson energy $E_J=¿$. On the other hand, when$\psi \vee \longrightarrow 0$the free energy $\overline{F}_1 \propto \psi \vee ❑^2|$. From these considerations, it is possible to arrive at differential equation governing the coupling energy (much similar to Josephson energy) of grains of any orientation individual (of its junctions) in the normal matrix [23]:

$$\nabla^2\vec{\gamma}=\frac{3}{\lambda_c^2}\frac{\vec{\gamma}\,d}{\gamma}\frac{d}{d\gamma}\left(1-\frac{\sin\gamma}{\gamma}\right) \qquad (42)$$

this equation describes the pseudo-phase $\gamma$,

$$\gamma=a) \qquad (43)$$

where, $\psi=\langle e^{i\phi}\rangle \longrightarrow \vee \psi \vee \langle e^{i\tilde{\phi}}\rangle$is the order parameter that determines any correlation between the individual phasesof any of the grains in the matrix. The magnetic field$\vec{B}(\vec{r})=\vec{\nabla}\times\vec{A}$in the sample results in an averagemagnetic fieldover all the elements of volumes (which were selected above in the definition of$\vec{h}$) containingmany grains. Thus, it was calculated the expression for$\lambda_c$ [23]:

$$\lambda_c=¿\ (44)$$

here, $z$ is the number of first neighbors of any grain, $I_J$is the critical chain of individual contacts of junctions. Thus, the Josephson effect (cooperative) that is dominant in coherence properties can be observed in the critical super current density and, more particularly,in its modulation by the external magnetic fieldapplied $H_{appl}^{dc}$. And, from the resolution of the differential equation 42for the pseudo-phase$\gamma$, we can get when $H<H_1(H_1=H_{c1})$ the following equation for the critical current, modulated by the applied magnetic field$H_{appl}^{dc}$:

$$I_c(H)=I_c(0)=0.82\left(1-\frac{1}{\gamma_s}\sin\gamma_s\right)^{1/2}\frac{H}{H_1}\ (45)$$

where $H_1$ it's the maximum value of the magnetic fieldwhere the Meissner effect occurs in a sample considered semi-infinite and $\gamma_s$it's the value of $\gamma$on the sample surface, and was estimated as:

$$\gamma_s\approx 2.7\left(\frac{R}{\lambda_c}\right)\left(\frac{H}{H_1}\right) \qquad (46)$$

where $R$ is the characteristic size of the sample. Thus, with these ideas and with the aid of the phase diagramat Ref. [XX] obtained through of our inductive measures (here we observed that our results have great reliability because, we can relate the inductive and magnetotransport measurements to explain our results without any change or mathematical adjustment) to the samplesfrom the system *Cuxwt%Nb*, we can understand and explain the results of diagram for the critical current $I_c(H)$in Ref. [XX].

Let's compare our results with the model discussed above for verify its validity. We started our discussion phase diagram obtained for the sample*Cu20wt%Nb* shown in Ref. [XX]. This is a magnetic phase diagram$H \times T$, constructed for the sample Cu20wt%Nb from the $M \times T$ measurements. The two critical lines of this diagram $H_c$and$H_{cJ}$are clearly defined in the onset of the transition and in the transition from weak coupling, where all points were extracted $T_{cJ}^{MXT}$are shown in the inset top of the figure. In the lower inset we show the region of fields magnetic $H_{c1J} \lll H \leq H_{c2J}$,which delimits the region governed by weak coupling between Nb grains. As was measured andshown previously in the other phase diagrams for the sample Cu20wt%Nb, $H_{c1J} = H^{\square}$. It is important to note that in the $H \times T$ diagram of Ref. [XX] the line $H_{c2}$is defined in the onset of the transition, and this line is associated exclusively to superconductive behavior volumetric ratio of the grains of Nb. As shown in the $M \times T$measurements, the critical temperature of this sample (Cu20wt%Nb) é$T_c^{Cu20wt\%Nb} = 8,2\,K$(H) and we know the reason of this value is lower than the expected transition to Nb, which iswith $T_{c-Bulk}^{(Nb)} = (9.25 \pm 2.02)\,K$[1]. All measurements $M \times T$, as shown in the figures 1 e 2, have a well-defined transition in $T_{cJ}^{MXT}$ in the ZFC(Zero Field Cooling measurement protocol at the PPMS® and MPMS® platforms)part for smaller magnetic fields than 200 Oe. This behavior is associated with the abrupt loss of coherence between the grains of Nb, which has as consequence a sudden penetration of magnetic flux into the sample at as T approaches $T_c$. This line was obtained through several $T_{cJ}$points for magnetic fields smaller than$H_{appl}^{dc} = 200\,Oe . \wedge \,, at\, the , I \times V$ measurements series, figures 16-18, wasadopt very important criterion to extract the spectrum of the $I_c^{Cu20wt\%Nb} \times H$points, with a cut-off voltage of$V_{cut-off}^{Cu20wt\%Nb} = 10^{-7}\,V$. According we have seen in the diagrams $M \times H$, in the figures 7-13, the signature of the weakregime coupling is shown by the presence of the two lines critical, upper $H_{c2}$and a lower $H_{c2J}$, whose regions of existence are limited by $H_{c1}^{Cuxwt\%Nb} \leq H_{appl}^{uppercritical} \leq H_{c2}^{Cuxwt\%Nb}$ in the upper critical line (volumetric superconductorbehavior) and$H_{c1J}^{Cuxwt\%Nb} \leq H_{appl}^{lowercritical} \leq H_{c2J}^{Cuxwt\%Nb}$on the lower criticalline (granular superconducting behavior).

Shown in Ref. [XX]: Magnetic phase diagram $H \times T$ constructed for the sampleCu20wt%Nb from the $M \times T$ measurements. The Critical Lines $H_c$and$H_{cJ}$are defined in the onset of the transition. The weak coupling transition from where the points were extracted$T_{cJ}^{MXT}(H)$are shown in the upper inset. At the lower inset we show the region of magnetic fields $H_{c1J}^{Cuxwt\%Nb} \leq H_{appl}^{lowercritical} \leq H_{c2J}^{Cuxwt\%Nb}$, which delimits the region governed by weak coupling between Nb grains.

In the magneticphase diagram $M \times T$ of Ref. [XX],as can be seen $H^{\square} \approx H_{cJ}$, what is the largest magnetic fieldwhere the Meissner effect occurs (according to we discussed). The all critical fields of granular coupling, $H_{c1J}$and$H_{c2J}$are located below the critical fields of volumetric behavior $H_{c1}$ and $H_{c2}$. The magnetic flux starts to penetrate at the Nb grains when$H_{appl}^{lowercritical}$ extrapolates the right edge of inequality $H_{c1J}^{Cuxwt\%Nb} \leq H_{appl}^{lowercritical} \leq H_{c2J}^{Cuxwt\%Nb}$, as a consequence of the gradual frustration of the Josephson coupling between the Nb grains, and this results in a reduction of the Josephson coupling and the electrical current throughout of the volume of whole sample. In the figure 13 we show the measured value of $H_{c2J}$, that is $H_{c2J}$=199,25 Oe. In thediagramof Ref. [XX]we show the curve describing the oscillatory behavior of the critical current that was normalized in$I_c(0) = 156\,mA$. The diagram clearly shows (which are damped with increasing magnetic field), and tend to disappear completely when the applied magnetic fieldvalueis close $H_{c2J}$($H_{c2J}$=199,25 Oe). These oscillations in the critical current are strongly modulated by the applied external magnetic fieldas we can observe. In this curve, the solid line of the diagram represents the adjustment we make through the use of equation 45. Still in the diagram, in the inset, is shown one of the $M \times H$ curves performed in this work for this sample, where we show the superconductorsignproduced by the weak coupling between Nb grains, and the overall sign due to the behavior volumetric Nb. Using always the criterion in these $I \times V$ measurementsthe spectrum of $I_c^{Cu20wt\%Nb}$data points from the cut-off voltage c$V_{cut-off}^{Cu20wt\%Nb} = 10^{-7} V$. The geometric properties of the sample *Cu20wt%Nb*-11-02 (shown in the table 1) was made as a strip form $l \times w \times t = 0.4\,cm \times 0.20\,cm \times 0.01\,cm$ to magnetotransport measures, because, when more strip-shaped, so much the better for this type of measure. We tookcare toapply the magnetic fieldin directionsperpendicular to the passing electric current in the sample. In the experimental curve of critical currentof thefigure 26,the behavior of the oscillations can be very clearly observed, and models very realistically the results provided in equation 45.To plot the critical current curve, we use$H^{\square} H_{cJ} = 15\,Oe$, value measured in the curve$M \times H\,(T = 4\,K)$,as may be seen in the inset of the critical current phase diagram in Ref. [XX]. We used the relationship of equation 46?

$$\gamma_s \approx 2.7 \left( \frac{R}{\lambda_c} \right) \left( \frac{H}{H_1} \right)$$

to obtain the constant $\zeta$,

$$\zeta \approx \gamma_s H \approx 2.7 \left( \frac{R}{\lambda_j} \right) \left( \frac{H}{H_1} \right) \approx 0.065 \qquad (48)$$

this implies

$$\left( \frac{R}{\lambda_c} \right) = \left( \frac{R}{\lambda_j} \right) \approx 1.5$$

$L = 2R = 2w$, which is the width of the strip prepared with the sample *Cux20%N*b for measures of magnetotransport(see table 1), and remembering that equation 45,provides for modeling of the behavior of the criticalcurrent for $R$. As can be verified, this result $\left( \frac{R}{\lambda_c} \right) = \left( \frac{R}{\lambda_j} \right) \approx 1.5$is within a value considered to be typical forgranular superconductors (mainly based on Nb grain). In the case of the sample *Cux20%Nb* (table 1),

$$\frac{w}{2} = \frac{R}{2} = \frac{L}{2} \approx \frac{0.2}{2}\,cm \approx 0.1\,cm \qquad (50)$$

Thus

$$\frac{w}{\lambda_c} = \frac{R}{\lambda_c} = \frac{L}{\lambda_j} \approx 1.5 \qquad (51)$$

and from equation 51, we estimate $\lambda_j$,

$$\frac{L}{\lambda_j} \approx 1.5 \Longrightarrow \lambda_j = \frac{L}{1.5} \Longrightarrow \lambda_j \approx \frac{0.1}{1.5}\,cm \approx 0.06\,cm \qquad (52)$$

and, we rewrite equation 45

$$I_c(H)=I_c(0)=0.82\left(1-\frac{1}{\gamma_s}\sin\gamma_s\right)^{1/2}\frac{H}{H_1} \quad (53)$$

at the following way:

$$I_c(H)=I_c(0)=0.82\left[1-\frac{1}{\zeta H}\sin(\zeta H)\right]^{1/2}\frac{H}{H_1} \quad (54)$$

We can visualize the comparison of this curve to our data points at the diagram at thefigure26.The adjustment performed through equation 55 satisfactorily follows the behavior of our experimental measures. It is important to note that the for which the oscillations in $I_c(H)$ is much lower than the measured fieldvalue $H_{c1J}$($H_{c1J}$=99,93 Oe) coupled the Nb grains. And the value of the fieldwhere the oscillations dampen completely isexactly in $H_{c2J}$($H_{c2J}$=199,25 Oe). If we instead of using the exponent $1/2$ in the equation 55, and rewrite itwith the exponent 2, this way

$$I_c(H)=I_c(0)=0.82\left[1-\frac{1}{\zeta H}\sin(\zeta H)\right]^{2}\frac{H}{H_1} \quad (55)$$

we can see the oscillations of current with the magnetic fieldin greater amplitude, and this provides a clearer idea of the agreement of the model with our experimental results for the sample *Cu20wt%Nb*. The line in Ref. [XX] represents this with excellent agreement between the results. From this curve (dashed line for the exponent 2), we can estimate a period of oscillation in

$$H\approx 24\,Oe \quad (56)$$

If we assume a grain network with the sufficiently regular spacing (SEM images in Ref. [16]) and grains with a good size distribution (see additionalinformationcontaining statistical analysis of the sample) of the structural properties of the sample *Cu20wt%Nb*, we can start from

$$\phi= B\times a \quad (57)$$

and obtain for sample *Cu20wt%Nb*

$$\phi_0=2.0678\times 10^{-7}\,Gcm^2\approx 24\langle a\rangle 58)$$

where $\langle a\rangle$is the mean value of the effective cross-section which separates two grains of Nb in the sample matrixCu20wt%Nb.

Shown in Ref. [XX]: Diagram showing the critical current normalized for $I_c(0)=156\,mA$. This diagram shows oscillations in the current modulated by the magnetic applied externalfield with period $H_{peri}=24\,Oe$. The line represents the adjustment made through the use of equation 45. On the inset is shown one of the performed $M\times H$ where we show the combined signals produced by weak coupling (Josephson Coupling) between the grains and the overall signal due to the volumetric behavior of Nb.

From equation 58, we can estimate $\langle a\rangle$from these magneto transport measures as follows

$$\phi_0=2.0678\times 10^{-7}\,Gcm^2\approx 24\langle a\rangle \Longrightarrow \langle a\rangle\approx\frac{2.0678\times 10^{-7}\,Gcm^2}{24}\approx 8.6\times 10^{-9}\,cm^2 (59)$$

from the equation 59 we can to estimate the average effective distance separating the Nb matrix, for the purpose of comparison with the results obtained through scanning microscopy. If we approach the junction area that are in contact with the grains by a circle (which is in fact very close to the real, since the grains have spheroidal form), we will have the following expression for the mean diameter of these circles

$$\langle a \rangle \approx \pi \langle R \rangle^2 (60)$$

where $\langle R \rangle$is the median radius of these circles, then

$$\langle R \rangle \approx \sqrt{\frac{\langle a \rangle}{\pi}} \Longrightarrow \langle R \rangle \approx \sqrt{\frac{8.6 \times 10^{-9}}{\pi}} \approx 0.52368 \times 10^{-5}\, m (61)$$

remembering, $2\langle R \rangle = \langle d \rangle$(relationship between diameter and radius), we will have

$$\langle d \rangle = 2\langle R \rangle \approx 2\sqrt{\frac{\langle a \rangle}{\pi}} \Longrightarrow \langle R \rangle \approx 2 \times 0.52368 \times 10^{-5}\, m \approx 1.0475\, \mu m \ (62)$$

The value for the mean diameter of the joints separating the grains from Nb in Cu matrix $\langle d \rangle \approx 1.0475\, \mu m$obtained in equation 62, it's at excellent agreement with the results measured for this distance in the scanning electron microscopy experiments $\langle d_m \rangle \approx 1.20\, \mu m$.

## IV. CONCLUSION

We developed a simple, robust and completely reproducible route to prepare a set of very good granular superconducting properties of the system samples *Cuxwt%Nb* (x = 3, 5, 15, 20), with the focus on the stoichiometry*Cu10wt%Nb* and *Cu20wt%Nb* a SEM-JEOL$^{Ltd}$ was used to characterization of the structural properties. In order to improve even more the arrangement of Nb grains in the Cu matrix, we submit the system to a third step, after RF melting, which we define as an improvement of grain rearrangement in the matrix annealing. This step consisted of placing the system in a resistive raise the system close to Cu melting temperature $T_{raise}^{\langle Cu \rangle} = 1100^o$C (Cu melting temperature Cu $T_{melting}^{\langle Cu \rangle} = 1083^o$C) and leave the system remain in this state for 48 hours until by difference of density and performance of viscous (and diffusion) forces, the grains of Nb to reach the ideal dispersion configuration in the Cu matrix.We made inductive measures *dc* magnetization *vs* temperature and magnetic field ($M^{dc}[H,T,\theta]$ by MPMS®with SQUIDdetection @ $H_{ext}^{max}$=70 kOe and $H_{ext}^{max}$=50 kOe), *ac* susceptibility measures vs temperature and magneto transport measurements (magnetization$\chi^{ac}[T, H, h_{Exc}, \upsilon, \theta,$ magneto-transport measures,$\rho^{dc}[H,T,\theta]$, $R^{dc}[V,IH,T,\theta]$, $I[V,H,T,\theta]$ and curves $J_c$ $[H,T,I,\theta]$ with PPMS® @ $H_{ext}^{max}$=90 kOe). We used a micro Hall sensorwith 10 measuring transverse voltage bridges regularly distributed throughout the longer sensor length, with dimensions $5.0 \times 0.2 \times 0.01$ $mm^3$ to calibrate de transverse voltage and obtain the best accuracy at the real external applied magnetic field value. We investigate the superconductingbehavior, coherence and quantum fluctuations properties nano-Nb superconducting tridimensionalarray of Josephson Junctions (3D-JJ) inCu-normal matrix. The precipitates of Nb that we obtained with the use of these techniques have sizes, $\langle D_m \rangle \approx 1.0475\, \mu m$, and spacing $\langle d_m \rangle \approx 1.20\, \mu m$, spheroidal shape geometry,well behaved andvery good quality.At the critical current curve $I[H,T,I,\theta]$ we show the emergence of a central maximum for small magnetic fields is followed by an excerpt with small oscillations with period of $H^{perio}$=24 Oe. In the presence of higher magnetic field ($H_{app}$), magnetic flux gradients arise and coherence is destroyed and the phase of the order parameter becomes random by suppressing the oscillations. The superconducting properties of the *Cuxwt%Nb* system are manifested usually through two important transition temperatures$T_c(H)$ and$T_{cJ}$, with ($T_{c-inter}^{(Cux20\%Nb)}$= 8.2 K). We have found $H_{c1J}^{(Cux20\%Nb)}$=99.93 Oe, $H_{c2J}^{(Cux20\%Nb)}$=199.25 Oe, $H_{c1}^{(Cux20\%Nb)}$=400.03 Oe, $H_{c2}^{(Cux20\%Nb)}$H=15 kOe and $T_{c-inter}^{(Cux20\%Nb)}$= 8.2 K, $R^{Cu20wt\%Nb} \approx 1.56 \times 10^{-4}\,\Omega$for the highest resistance saturation value for $H_{appl}^{dc}$= 0 and $\theta=0^o\,\theta=45^o$and $\theta=90^o$. Also, we found values close to $\rho^{Cu20wt\%Nb} \approx 10^{-24}\,\Omega cm$ for the resistivity.

## V. FINAL REMARKS

Initially, we did not intend to relate the response of our samples to an oscillating fieldwith current transport measures. This is because, because of the possibility of a frustration of the coupling of the amplitudes and oscillation acoustic phase of the wave functions of the Cooper pairs formed in the individual grains of the matrix corresponds to the sample. This behavior explains why in the literature, we use the term "*weak coupling* "or “weak links” to this type of superconducting response, and therefore, even though the susceptibility measures are a good tool for extracting the critical current from materials mainly superconducting materials, bulk materials, we initially considered the possibility of using such measures extrapolate results and get some information from sample transport properties Cuxwt%Nb. However, after optimizing some measurement parameters in our experimental set up system, mainly the Hall sensor for the calibration and greater control of the magnetic fieldvalues in the sample region, we realized that it was possible to use the magnetotransport and correlates them with the magnetic susceptibility in our granular system. We studied a theoretical model at the references [23] describing effects of coherence and quantum interference in the presence of low fields and low electrical currents that we consider to be quite complete and test it in our system. This model is well suited to explain the experimental details of our results quantitative and qualitative characteristics in the granular superconducting system *Cuxwt%Nb*. We correlate the properties structural, magnetic and transport of our samples and found satisfactory results. We checked experimentally and explain the existence of the central maximum followed by the long tail in the critical current curve $I_c$[H] or [H,T,I,$\theta$])and the occurrence of small oscillations due to the effects of quantum interference.

## SUPPLEMENTARY MATERIAL

Albino verá que o texto ficou extenso mas bastante robusto. De fácil leitura (deixei bem claro e ditático). Tem muita física interessante aqui. Gosto muito desse trabalho. É simples e com muita informação científica para a supercondutividade. Aqui nessa seção, conforme mencionei no texto, terei que colocar as curvas de EDS e EDX sob forma de informação suplementar. O texto já está extenso e não acho razoável colocar essas curvas. De fato, elas não são necessárias aqui. Aguardo sua opnião. Ou menciono de forma suplementar ou não menciono, deixo a critério da solicitação dos referees.

## ACKNOWLEDGEMENTS

We acknowledge the support from Brazilian science agencies CAPES, CNPq and FAPESP for financial support.